\documentclass[a4paper,11pt]{article}
\pdfoutput=1
\usepackage{jheppub}
\usepackage[utf8x]{inputenc}
\usepackage{amsmath}
\usepackage{latexsym}
\usepackage{amssymb}
\usepackage[english]{babel}
\usepackage{color}
\usepackage[multiple]{footmisc}
\usepackage{breakurl}
\usepackage{mathrsfs}
\usepackage{amsmath}
\usepackage{amsfonts}
\usepackage{mathrsfs}
\usepackage{slashed}
\usepackage{epigraph}
\usepackage{amssymb}
\usepackage{pifont}
\usepackage{fancyhdr}
\usepackage{amssymb}
\usepackage{graphicx}
\usepackage{longtable}
\usepackage{makecell}
\usepackage{amscd}
\usepackage{tabu}
\usepackage{bm}
\usepackage{enumerate}
\usepackage{caption}
\usepackage{subcaption}
\usepackage{color}

\usepackage{hyperref} 
\hypersetup{linktocpage=true}
\usepackage[all]{hypcap}
\hypersetup{
   bookmarks=true,         
   unicode=false,          
   pdftoolbar=true,        
   pdfmenubar=true,        
   pdffitwindow=false,     
   pdfstartview={FitH},    
   pdftitle={My title},    
   pdfauthor={Author},     
   pdfsubject={Subject},   
   pdfcreator={Creator},   
   pdfproducer={Producer}, 
   pdfkeywords={keyword1} {key2} {key3}, 
   pdfnewwindow=true,      
   colorlinks=true,        
   linkcolor=blue,         
   citecolor=magenta,      
   filecolor=magenta,      
   urlcolor=cyan           
}
\usepackage{fancyhdr}
\usepackage{booktabs}
\definecolor{verdes}{cmyk}{0.92,0,0.59,0.4}

\newcommand{\bea}{\begin{eqnarray}}
\newcommand{\eea}{\end{eqnarray}}

\preprint{Cavendish-HEP-15/08, DAMTP-2015-56}
\title{Linear flavour violation and anomalies in $B$ physics}
\author[a]{Ben Gripaios,}
\author[a,b]{Marco Nardecchia,}
\author[b]{Sophie Renner}

\affiliation[a]{\normalfont{Cavendish Laboratory, University of Cambridge, 
J.J. Thomson Avenue, Cambridge, CB3 0HE, UK}}
\affiliation[b]{\normalfont{DAMTP, University of Cambridge, 
Wilberforce Road, Cambridge, CB3 0WA, UK}}

\emailAdd{gripaios@hep.phy.cam.ac.uk} 
\emailAdd{m.nardecchia@damtp.cam.ac.uk}
\emailAdd{sar67@cam.ac.uk} 

\abstract{We propose renormalizable models of new physics that can explain various anomalies observed in decays of $B$-mesons to electron and muon pairs. 
The new physics states couple to linear combinations of Standard Model fermions, yielding a pattern of flavour violation that gives a consistent fit to the gamut of flavour data. Accidental symmetries prevent contributions to baryon- and lepton-number-violating processes, as well as enforcing a loop suppression of new physics contributions to flavour violating processes. 
Data require that the new flavour-breaking couplings are largely aligned with the Yukawa couplings of the SM and so we also explore patterns of flavour symmetry breaking giving rise to this structure.}

\begin{document}
\maketitle

\section{Introduction}
Over the last few decades, the Standard Model (SM) has emerged victorious as the correct description of physics up to and somewhat beyond the weak scale. Nowhere is this more true than in the flavour sector, where (with the addition of neutrino masses) the Yukawa interactions give rise to a pattern of fermion masses, mixings, and CP violation whose consistency with myriad precise experimental observations is striking.
Crucial to this consistency is the particular (even peculiar) structure of the SM, which leads to many subtle and delicate cancellations in flavour-changing and CP-violating processes. As examples at tree level (see, {\em e.g.\@} \cite{Gripaios:2015gxa} for more details), the representation content of the SM fermions (which is such that any two equivalent irreducible representations under the unbroken $SU(3)_c \times U(1)_{em}$ are also equivalent under the full $SU(3)_c \times SU(2)_L \times U(1)_Y$)
ensures the absence of flavour-changing neutral currents (FCNCs) mediated by vector bosons, while the presence of just a single Higgs doublet field ensures the absence of FCNCs mediated by the Higgs boson. At loop level, the unitarity of the CKM matrix, together with the fact that it is empirically observed to be close to being diagonal, leads to remarkable suppressions of flavour-changing processes via the GIM mechanism. 

Key to all of these suppressions, of course, is our insistence that the SM be renormalizable; indeed, if we relax this requirement, we quickly run into conflict with data, unless the UV cut-off of the theory is rather high. 

In the last year or so, cracks have begun to appear in the SM edifice, in the form of a variety of anomalies associated with $B$-meson decays. These include effects seen in angular observables \cite{Matias:2012xw,DescotesGenon:2012zf,Descotes-Genon:2013vna,Altmannshofer:2013foa,Descotes-Genon:2014uoa,Hurth:2014vma,Altmannshofer:2014rta,Descotes-Genon:2015xqa,Descotes-Genon:2015uva} of the decay $B \to K^* \mu \mu$ \cite{Aaij:2013qta}, the observable $R_K$ \cite{Aaij:2014ora}, and a series of branching ratios with muons in the final state (\textit{e.g.\@} \cite{Aaij:2014pli,Aaij:2015esa}). Whilst it is surely far too soon to claim that the end of the SM is nigh, it is perhaps worthwhile to explore how the SM might be modified in such a way that these anomalies in the data can be accommodated. Of course, this must be done in such a way as not to spoil the various delicate cancellations that occur in the SM, since this would lead to gross contradictions with data elsewhere. As such, it makes sense to look for theory which is renormalizable (or has a cut-off which is much larger than the scale of new physics, which amounts to the same thing).
Then we can at least hope that, just like in the SM, dangerous flavour-changing processes can be kept under control. 

A number of models explaining the anomalies have already been proposed. These can be sub-divided into models which generate the anomalies via tree- {\em vs.} loop-level corrections. Models in the first category include leptoquarks \cite{Hiller:2014yaa,Biswas:2014gga,Gripaios:2014tna,Sahoo:2015wya,Varzielas:2015iva,Becirevic:2015asa,Alonso:2015sja,Calibbi:2015kma,Sahoo:2015qha}
 and $Z^\prime$s \cite{Descotes-Genon:2013wba,Gauld:2013qba,Buras:2013dea,Buras:2013qja,Altmannshofer:2014cfa,Buras:2014yna,Crivellin:2015mga,Crivellin:2015lwa,Niehoff:2015bfa,Sierra:2015fma,Crivellin:2015era,Celis:2015ara,Greljo:2015mma,Niehoff:2015iaa,Altmannshofer:2015mqa,Falkowski:2015zwa}. Not all of these models are renormalizable. Ref.\ \cite{Gripaios:2014tna}, for example, describes a well-motivated model in which the electroweak hierarchy problem is solved by Higgs compositeness and SM fermions are partially composite. A light leptoquark can arise naturally as a Goldstone boson \cite{Gripaios:2009dq}. But if the compositeness scale is sufficiently high, the renormalizable limit is approximately recovered. There is one recently-proposed model in the second category, which generates contributions that explain the $B$ anomalies and the anomalous magnetic moment of the muon, via a $Z^{\prime}$ with loop-induced couplings to muons \cite{Belanger:2015nma}.

In this work, we wish to propose a second type of renormalizable model in the category of those generating the anomalies at loop level. In order to obtain the flavour structure that seems to be required (in the quark sector at least) in an automatic way, we insist that the new flavour-violating couplings be linear in the SM fermions. We then survey the various possibilities for the BSM fields and find that basic phenomenological considerations (such as the insistence on an accidental symmetry stabilising the proton), together with a minimality criterion, lead to just two possible models. The models both feature two new scalar fields and a single fermion field, which couple to linear combinations of the SM fermions via Yukawa interactions. The two models are very similar in their phenomenology and so we discuss only one in detail. We find that, for suitable values of the parameters, a satisfactory fit to the anomalies can be obtained.

The fit to the data requires that the new flavour-violating couplings be strongly hierarchical (at least in the quark sector; in the lepton sector there is more room to manoeuvre) and moreover largely aligned with the flavour breaking already present in the Yukawa couplings of the SM. Thus, even more than in the SM, the low-energy theory seems to be crying out for a fundamental theory of flavour.
Rather than attempt to find an explicit theory of flavour that does the job, we content ourselves with showing the plausibility of the existence of such a theory, by exhibiting patterns of flavour symmetry breaking that can give rise to the required structure. We find that a variety of symmetries are possible. Interestingly, they give rise to a pattern of couplings similar to that obtained in theories featuring partial compositeness.

The models have another feature of interest. By construction, they yield explicit UV completions that generate the non-renormalizable flavour structure that was identified in \cite{Glashow:2014iga} as a viable one for explaining the anomalies. The fact that such UV completions exist is a desirable thing, because the alignment of the flavour-violating new physics couplings in \cite{Glashow:2014iga} is not preserved by the SM RG flow. 
Thus, in the absence of an unexplained fine tuning in the couplings, the scale of new physics completing the effective lagrangian in \cite{Glashow:2014iga} should be light, in order that the picture makes sense. Our models provide explicit completions.

The structure of the paper is as follows. In Section \ref{model} we describe the construction and content of the models. In Section \ref{pheno} we outline phenomenological implications of one of the models and find allowed parameter space regions to fit the $B$ anomalies. Section \ref{flavour} contains an analysis of possible flavour breaking patterns, and our conclusions are given in Section \ref{conclusions}.

\section{The Models}
\label{model}

Recent experimental data in flavour physics, in particular measurements by the LHCb collaboration \cite{Aaij:2013qta,Aaij:2014ora,Aaij:2015esa}, suggest possible effects of New Physics (NP) in semileptonic decays of $B$-mesons. In particular, considering the $b \to s$ quark flavour transitions, the most significant departures from the Standard Model predictions are observed in: \textit{(i)} the so-called $P'_5$ angular observable of the $B \to K^* \mu^+ \mu^-$ decays \cite{Aaij:2013qta} \textit{(ii)} a series of branching ratios of B-decays with muons in the final states (like the recently measured $B^0_s \to \phi \mu^+ \mu^-$ \cite{Aaij:2015esa}) \textit{(iii)} the observable $R_K$ \cite{Aaij:2014ora} defined as the ratio of branching ratios of $B \to K \mu^+ \mu^-$ and $B \to K e^+ e^-$ in the low $q^2$ (lepton pair invariant mass squared) region. It is perhaps premature to claim evidence for NP. Indeed the discrepancies in \textit{(i)} and  \textit{(ii)} could be due to underestimates of hadronic uncertainties\footnote{For discussions on the size of the hadronic uncertainties on these observables we refer the reader to {\em e.\@g.} \cite{Khodjamirian:2010vf,Jager:2012uw,Beaujean:2013soa,Descotes-Genon:2014uoa,Lyon:2014hpa,Descotes-Genon:2014joa,Altmannshofer:2014rta,Jager:2014rwa,Altmannshofer:2015sma,Descotes-Genon:2015uva}.}, while the discrepancy in \textit{(iii)}, despite the observable being theoretically very clean in the SM, could simply be a statistical fluctuation. 
However, if we allow for an interpretation of these experimental results in terms of NP, it is quite remarkable that model-independent approaches based on higher-dimension operators \cite{Descotes-Genon:2013wba,Altmannshofer:2013foa,Beaujean:2013soa,Horgan:2013pva,Datta:2013kja,Hurth:2013ssa,Alonso:2014csa,Hiller:2014yaa,Ghosh:2014awa,Hurth:2014vma,Altmannshofer:2014rta,Altmannshofer:2015sma} give a simple and consistent fit to the data.

In general, we expect that NP will couple chirally to the matter fields; assuming a coupling purely to either right- or left-handed
currents, the fits find that the operator $\overline{b}_L \gamma^{\alpha} s_L \, \overline{\mu}_L \gamma_{\alpha} \mu_L$ is preferred.
Motivated by this, we seek renormalizable models that couple the NP to quark and lepton doublets. Moreover, for reasons that will become clear, we demand that the SM fermions appear linearly in the NP couplings. In what follows, the SM fermions are written as $Q_L, U_R,D_R,L_L,E_R$ while the beyond SM (BSM) fermions and scalars are denoted by $\Psi_i$ and $\Phi_i$.
The Higgs doublet is denoted by $H$ and transforms as $(1,2,\tfrac{1}{2})$. The possible linear interactions of the new BSM fields with the SM doublets ($Q_L^i$ and $L_L^i$) can be classified in the following way:
\begin{align}
\label{eq:yukone}
\alpha^q_i \, &\overline{\Psi} Q_L^i \Phi_q + \alpha^{\ell}_i \, \overline{\Psi} L_L^i  \Phi_{\ell} + \mathrm{h.c.}; \\
\label{eq:yuktwo}
\mathrm{or} \;\; \alpha^q_i \, &\overline{\Psi} Q_L^i  \Phi_q + \alpha^{\ell}_i \,  \overline{\Psi}^c  L_L^i \Phi_{\ell} + \mathrm{h.c.};\\ 
\mathrm{or} \;\;\alpha^q_i \, &\overline{\Psi}_q Q_L^i \Phi + \alpha^{\ell}_i \, \overline{\Psi}_{\ell} L_L^i \Phi + \mathrm{h.c.}; \\
\mathrm{or} \;\;\alpha^q_i \, &\overline{\Psi}_q Q_L^i  \Phi + \alpha^{\ell}_i \,  \overline{\Psi}_{\ell}^c L_L^i \Phi + \mathrm{h.c.} 
\end{align}
Notice that one new state appears simultaneously in both the quark and lepton interactions (this is needed in order to draw a diagram contributing to $b \to s \mu \mu$). In the first two cases the common mediator is $\Psi$ while in the last two the common mediator is $\Phi$. There are infinitely many combinations with suitable gauge quantum numbers to yield these interactions. In order to reduce the possibilities for the SM quantum numbers of the new states, we will impose conditions on them based on the following considerations:

\begin{enumerate}[(a)]
\item Accidental protection

The irreducible representations (`irreps') of new states should be such that other renormalizable couplings with SM fermions beyond the Yukawa terms above are forbidden by gauge invariance. This guarantees that the definitions of Baryon and Lepton number can be extended to include the new sector, and that they will then remain accidental symmetries of the model. It also prevents other sources of flavour breaking. This criterion immediately implies that none of the new states can be in the same representations under the SM gauge and Lorentz groups as any SM field. 

\item Scalar couplings with the Higgs doublet

New scalars will always have quartic interactions with the Higgs doublet. In particular, terms of the form $(\Phi^{\dagger} \Phi) (H^{\dagger} H)$ and $(\Phi^{\dagger} T^a_{\Phi} \Phi) (H^{\dagger} T^a_{H} H)$ are never forbidden by gauge symmetry (though the second one is absent if $\Phi$ is a $SU(2)_L$ singlet). These terms are phenomenologically viable, but other quartic and trilinear interactions of a scalar with the Higgs, like $\Phi H H$, or $\Phi H H H$, could give rise to a violation of the custodial symmetry at the tree level and/or could modify the observed Higgs phenomenology. To be safe from these unwanted effects, we choose the quantum numbers of the new scalars such that these dangerous interactions are prohibited.

\item Direct searches, coloured particles

New particles in the loops will need to be rather light to create a measurable effect in $B$ decays, so it is convenient to choose quantum numbers such that their masses are less constrained by direct searches. The quark interaction requires at least one state that transforms non-trivially under the colour group. A coloured scalar will have weaker bounds on its mass than a coloured fermion, since in the latter case the production cross section is higher for a fixed gauge quantum number. 
Selecting a scalar to be the only new coloured particle leads us to consider only the first two cases in the above list of new interactions, namely the ones with a single fermion mediator $\Psi$ and two different scalars $\Phi_q$ and $\Phi_{\ell}$.

\item Direct searches, BSM Lightest Particle (LP)

The Yukawa interactions above are manifestly invariant under a $U(1)$ transformation that acts non trivially only on the BSM states. This symmetry is respected by the gauge-kinetic terms of the new states too. 
We look for irreps such that this transformation is an accidental symmetry of the whole renormalizable model. This has the advantage that all NP flavour-violating processes are loop suppressed. But it also implies that the lightest NP state is stable; in order to evade strong constraints on coloured and/or electrically charged stable states coming from colliders and cosmology we look for a colourless irrep containing a neutral particle. The gauge quantum numbers are $(1,n,y)$ under $SU(3)_C \times SU(2)_L \times U(1)_Y$ and requiring an uncharged lightest state gives the restriction that $-n/2 \leq y \leq n/2$ w $y$ is integer if $n$ is odd and is half-integer if $n$ is even.

\item $SU(2) _L$ Minimality 

Finally we apply a minimality criterion on the dimensionality of the SU(2) irreps. This criterion is harder to justify on phenomenological grounds, though larger representations will, of course, lead to a Landau pole at lower energies. For the sake of argument, we require that all new irreps have dimensionality fewer than $5$.

\end{enumerate}

Let us now identify the irreps that could contain the LP. From (d) and (e) we obtain a finite list of candidates. Most of them are excluded because of the presence of renormalizable interactions that violate conditions (a) and/or (b). The excluded cases are listed in Tab.\@ \ref{irreps}. 
\begin{table}
\begin{equation*}
\begin{array}{c|c|c}
\textrm{Field} & SU(3)_C \times SU(2)_L \times U(1)_Y & \textrm{Interactions} \\
\hline
\hline
\Phi_{\ell} & (1,1,0) &   \overline{\Psi} L^i_L \textrm{ or } \overline{\Psi}^c L^i_L\\
 & (1,2,\tfrac{1}{2}) & \Phi_{\ell} \overline{U}^i_R Q^j_L  \\
 & (1,3,0) & \Phi_{\ell} \, H^{\dagger} H\\
& (1,3,1) & \Phi_{\ell} \, H^{\dagger} H^{\dagger} \\
& (1,4,\tfrac{1}{2}) & \Phi_{\ell} \, H^{\dagger} H^{\dagger} H \\
& (1,4,\tfrac{3}{2}) &  \Phi_{\ell} \, H^{\dagger} H^{\dagger} H^{\dagger} \\
\hline
\Psi & (1,1,0) & \overline{\Psi} L^i_L H \\
 & (1,2,\tfrac{1}{2}) & \overline{\Psi} L^i_L \\
& (1,3,0) & \overline{\Psi} L^i_L H\\
& (1,3,1) & \overline{\Psi}^c L^i_L H^{\dagger} \\
\hline
\end{array}
\end{equation*}
\caption{Irreducible representations (with $d<5$) of multiplets that contain a colourless and neutral particle but which are rejected because they could give rise to unwanted renormalizable interactions, as listed in the last column. Irreps with negative hypercharges are related to these ones by charge conjugation.}
\label{irreps}
\end{table}
We are left with four cases, each of which has a single fermion, $\Psi$, with SM quantum numbers $(1,4, \pm 1/2)$ or $(1,4, \pm 3/2)$. However, radiative corrections split the values of the particle masses in the multiplet and it turns out that, for the quantum number $(1,4, \pm 1/2)$, the LP is not the neutral one.
We conclude that, since we are demanding a neutral LP, the LP can only be contained in the fermion field $\Psi$ with quantum numbers $(1,4, \pm \tfrac{3}{2})$. Imposing condition (e) on the field $\Phi_q$ we are left with just two models:
\begin{itemize}
\item Model A. $\Psi \sim (1,4,+\tfrac{3}{2}), \Phi_q \sim (\overline{3},3,\tfrac{4}{3}) $,  $\Phi_{\ell} \sim (1,3,2) $
with Yukawa interactions as in (\ref{eq:yukone}):
\begin{equation}
\label{ModelA}
\alpha^q_i \, \overline{\Psi} Q_L^i \Phi_q + \alpha^{\ell}_i \, \overline{\Psi} L_L^i  \Phi_{\ell} + \textrm{h.c.} 
\end{equation}
\item Model B. $\Psi \sim (1,4,-\tfrac{3}{2}), \Phi_q \sim (\overline{3},3,-\tfrac{5}{3}) $,  $\Phi_{\ell} \sim (1,3,2) $
with Yukawa interactions as in (\ref{eq:yuktwo}):
\begin{equation}
\alpha^q_i \, \overline{\Psi} Q_L^i  \Phi_q + \alpha^{\ell}_i \,  \overline{\Psi}^c  L_L^i \Phi_{\ell} +\textrm{h.c.} 
\end{equation}
\end{itemize}
The two models have very similar implications for the phenomenology that we are interested in here. Henceforth, we discuss only Model A.

The quantum numbers of the SM and NP fields under the gauge and global symmetries (to be discussed below) are summarised in Tab.\@ \ref{quantumnumbers}
\begin{table}
\begin{equation*}
\begin{array}{c|c|c}
\textrm{Field} & SU(3)_C \times SU(2)_L \times U(1)_Y & U(1)_{B'} \times U(1)_{L'} \times U(1)_{\chi} \\
\hline
\hline
Q_L & (3,2,\tfrac{1}{6}) & (\tfrac{1}{3},0,0) \\
U_R & (3,1,\tfrac{2}{3}) & (\tfrac{1}{3},0,0) \\
D_R & (3,1,-\tfrac{1}{3}) & (\tfrac{1}{3},0,0) \\
L_L & (1,2,-\tfrac{1}{2}) & (0,1,0) \\
E_R & (1,1,-1) & (0,1,0) \\
\Phi_H & (1,2,\tfrac{1}{2}) & (0,0,0) \\
\hline
\Psi & (1,4,-\tfrac{3}{2}) & (0,0,1) \\
\Phi_q & (\overline{3},3,\tfrac{4}{3}) & (-\tfrac{1}{3},0,1) \\
\Phi_ {\ell} &(1,3,2) & (0,-1,1) \\
\hline
\end{array}
\end{equation*}
\caption{Quantum numbers of the Standard Model fields and new fields under the SM gauge symmetry (second column), and under the accidental global symmetries of the theory (third column).}
\label{quantumnumbers}
\end{table}
and the most general renormalizable lagrangian is given by
\begin{eqnarray}
\mathcal{L} &=& \mathcal{L}_{SM} + \mathcal{L}_{\Phi} + \mathcal{L}_{\Psi} + \mathcal{L}_{\textrm{yuk}}, \\
 \mathcal{L}_{\Phi} &=& \left(D^{\mu} \Phi_{\ell}\right)^{\dagger} D^{\mu} \Phi_{\ell} 
+\left(D^{\mu} \Phi_{q}\right)^{\dagger} D^{\mu} \Phi_{q} - V(\Phi_H,\Phi_q,\Phi_{\ell}), \\
\mathcal{L}_{\Psi} &=& i \overline{\Psi} D^{\mu} \gamma_{\mu} \Psi - M_{\Psi} \overline{\Psi} \Psi , \\
\label{lin}
\mathcal{L}_{\textrm{lin}} &=& \alpha^q_i \, \overline{\Psi}_R Q^i_L \Phi_q + \alpha^{\ell}_i \, \overline{\Psi}_R L^i_L \Phi_{\ell} +
\alpha^{q *}_i \, \overline{Q}^i_L \Psi_R \Phi^{\dagger}_{q} + \alpha^{\ell *}_i \, \overline{L}^i_L \Psi_R \Phi^{\dagger}_{\ell} .
\end{eqnarray}
See Appendix \ref{SUtwodecomp} for the explicit decompositions of the operators in terms of components of the $SU(2)_L$ multiplets.
Let us now analyse the accidental global symmetries of this lagrangian. Before considering the breaking coming from $\mathcal{L}_{\textrm{lin}}$ it is easy to show that the Lagrangian is invariant under a global $U(1)^7$. Indeed, the SM alone has accidental global symmetry $U(1)_B \times U(1)_e \times U(1)_{\mu} \times U(1)_{\tau} $, while the gauge kinetic terms of the new BSM fields have global symmetry  $U(1)_{\Psi} \times U(1)_{\Phi_{q}} \times U(1)_{\Phi_{\ell}}$. Moreover, it is easy to prove that the most general renormalizable scalar potential $V(\Phi_H,\Phi_q,\Phi_{\ell})$ is invariant under $U(1)^7$. 

Now consider the effect of $\mathcal{L}_{\textrm{lin}}$. For a generic choice of the couplings $\alpha^{\ell}$ and $\alpha^q$ there is always an unbroken $U(1)^3 \equiv U(1)_{B'} \times U(1)_{L'} \times U(1)_{\chi}$, defined as follows. 
Under the $U(1)_{B'}$ the SM fields have their usual baryon number while $\Phi_q$ has charge -1/3. Similarly,
under the $U(1)_{L'}$ the SM fields have their usual lepton number while $\Phi_{\ell}$ has charge -1.
Finally the SM fields are uncharged under $U(1)_{\chi}$, while the BSM fields have charge unity\footnote{This symmetry makes our neutral and colourless LP stable. Hence, the LP is a potential dark matter (DM) candidate. However, even if its mass and couplings could be fixed in order to reproduce the right relic abundance, data from direct detection experiments exclude it as DM, because of its large coupling to the Z-boson. A potentially large relic density could be problematic, though non-renormalizable operators are expected to break the accidental $U(1)_{\chi}$ at very high scales, triggering its decay into SM particles.}.

Thus, the model retains analogues of the accidental baryon and lepton number symmetries of the SM, which suffice to stabilize the proton and to prevent contributions to numerous unobserved lepton- and baryon-number violating processes. Moreover, the model features an additional accidental $U(1)_{\chi}$ symmetry, under which SM fields are uncharged. An immediate consequence of this is that all NP-generated processes involving only SM particles in the initial and final states are loop-suppressed. This is certainly an advantage from the point of view of the vast majority of flavour-violating observables, where no deviation from the SM is observed. It might be regarded as a disadvantage from the point of the view of the $B$-physics anomalies, where a sizable NP effect is needed. But this is offset somewhat by the desirable structure of linear NP flavour violation that results. We shall see in the sequel that the anomalies can be reproduced even for values of the NP couplings that are of order unity or smaller.

\section{Phenomenological Analysis}
\label{pheno}
In this section we discuss the phenomenology of 
Model A. In an obvious notation, we denote the masses of the new states as $M_{\Psi}$, $M_q$, and $M_{\ell}$. In a basis where the left-handed quark doublet is defined as $Q^i_L = (V_{CKM}u^i_L,d^i_L)^T$, the minimal set of couplings $\alpha$ that are needed to fit the $b \to s \ell \ell$ anomalies are $\alpha_3^q$, $\alpha_2^q$ and $\alpha_2^{\ell}$ (\emph{i.e.} couplings involving $b$, $s$ and $\mu$). To begin with, we will assume that only these couplings are non-zero and investigate the processes induced. In this section we collect relevant formula on indirect searches, and investigate direct production bounds. In subsection \ref{plots} we use this information to find allowed parameter space regions.
 In section \ref{flavour} we will discuss relaxing the assumptions on the couplings and propose more motivated flavour structures.

\subsection{Indirect searches}

As described above, the accidental global $U(1)_{\chi}$ symmetry under which the new particles are charged implies that contributions to processes containing only SM particles in the initial and final states are only induced at loop level. Here we investigate the size of these contributions. 

The relevant processes, given the assumption on couplings described above, are $b \to s \mu \mu$ processes, $B_s$ mixing, $b \to s \gamma$, and the anomalous magnetic moment of the muon.

\subsubsection{Semileptonic four-fermion operators}
\label{bsllsec}
\begin{figure}
\begin{center}
\includegraphics[width=5.5cm]{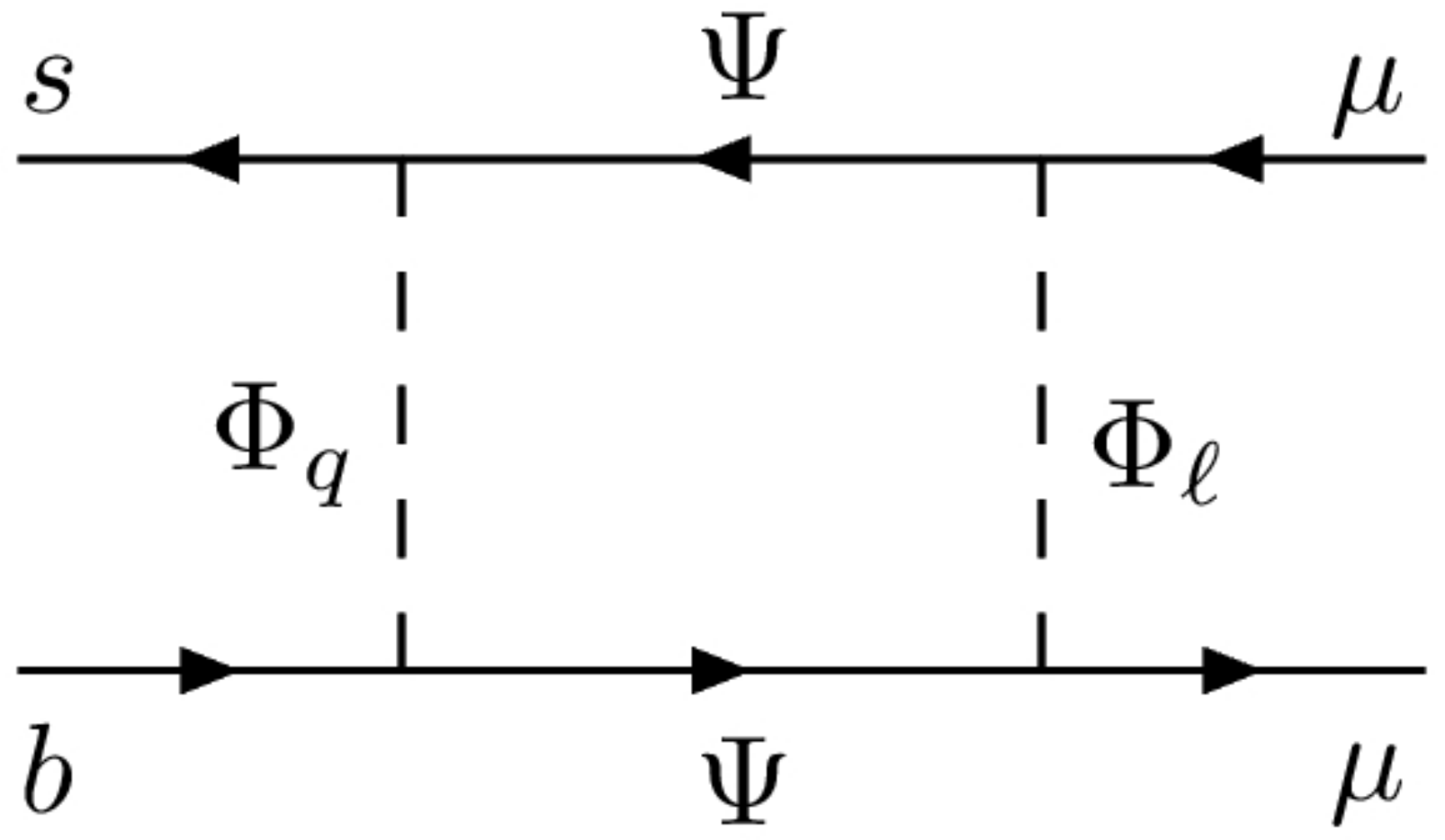}
\end{center}
\caption{Diagram contributing to $b \to s \mu \mu$}
\label{bsll}
\end{figure}

The process $b \to s \ell \ell$, important for the LHCb $B$ meson anomalies, is induced at loop level by the diagram in Fig.\@ \ref{bsll}\footnote{There are also $Z$ and photon penguin diagrams which contribute, with a NP loop connecting the quarks and joining to the leptons via a $Z/\gamma$ propagator. These penguin diagrams are discussed in Appendix \ref{penguins} and are found to be very suppressed relative to both the SM contribution and the diagram in Fig.\@ \ref{bsll}, and hence are neglected here.}. The $SU(2)_L$ structure of the NP-induced semileptonic four-fermion interaction can be derived from the discussion in Appendix~\ref{SUtwodecomp}, using the lagrangian (Eqn.\@ \ref{SU2decomplagrangian}) written explicitly in terms of $SU(2)_L$ components. The resulting effective NP lagrangian is

\begin{equation}
\mathcal{L}_{eff} \supset \frac{K({x_q,x_{\ell}})}{M^2_{\Psi}} \frac{\alpha^{q*}_i \alpha^{q}_j \alpha^{\ell*}_m \alpha^{\ell}_n}{64 \, \pi^2 } \left[ \left( \overline{Q}^i_L \gamma^{\mu} Q^j_L\right) \left( \overline{L}^m_L \gamma_{\mu} L^n_L \right) + \frac{5}{9} \left( \overline{Q}^i_L \gamma^{\mu} \vec{\tau} Q^j_L\right) \cdot \left( \overline{L}^m_L \gamma_{\mu} \vec{\tau} L^n_L \right) \right],
\label{bsllNPeffL}
\end{equation}
with $x_{q} \equiv \frac{M^2_{q}}{M^2_{\Psi}}$ and $x_{\ell} \equiv \frac{M^2_{\ell}}{M^2_{\Psi}}$. The loop function $K({x_q,x_{\ell}})$ can be obtained by the following definitions;
\begin{eqnarray*}
K(x) & \equiv & \frac{1-x+x^2 \log x}{(x-1)^2}, \\
K(x,y)& \equiv & \frac{K(x)-K(y)}{x-y}.
\end{eqnarray*}
The effective hamiltonian relevant to $b \to s \ell \ell$ transitions is

\begin{equation}
{\mathcal H}_{\rm eff} = - 
\frac{4 G_F}{\sqrt{2}}\, (V_{ts}^\ast V_{tb}) \,  
\sum_{i}^{} C^{\ell}_i (\mu) \, {\mathcal O}^{\ell}_i (\mu) \, \, ,
\label{effH}
\end{equation}
where $\mathcal{O}^{\ell}_i$ are a basis of $\rm SU(3)_C \times U(1)_Q$-invariant 
dimension-six operators giving rise to the flavour-changing transition. The superscript $\ell$ denotes the lepton flavour in the 
final state $(\ell \in \{e,\mu,\tau\})$, and the important operators for our process,
$\mathcal{O}^{\ell}_i$, are given in a standard basis by
\begin{eqnarray}
{\mathcal O}^{\ell (')}_9 &=& \frac{\alpha_{\rm em}}{4 \pi} \,  
\left (\bar s \gamma_\alpha P_{L(R)} b \right ) (\bar \ell \gamma^\alpha \ell) 
\; , \; \rm \label{WilsonOps} \\ 
{\mathcal O}^{\ell (')}_{10} &=& \frac{\alpha_{\rm em}}{4 \pi} \, 
\left (\bar s \gamma_\alpha P_{L(R)} b \right ) 
(\bar \ell \gamma^\alpha \gamma_5 \ell). \nonumber
\end{eqnarray}
Comparing equations \ref{bsllNPeffL} and \ref{effH} we find the NP contribution to the Wilson coefficients relevant to $b \to s \mu \mu$ is
\begin{equation}
\label{boxbsll}
C^{\mu NP}_9=-C^{\mu NP}_{10}= \left( \frac{4 G_F}{\sqrt{2}} V^*_{ts} V_{tb} \frac{\alpha}{4 \pi}\right)^{-1} \frac{7}{576 \pi^2} \frac{K({x_q,x_{\ell}})}{M^2_{\Psi}} \alpha^{q*}_2 \alpha^{q}_3 \left| \alpha^{\ell}_2 \right|^2.
\end{equation}
The most recent best fit ranges on this combination of Wilson coefficients are taken from \cite{Altmannshofer:2015sma} and are given by
\begin{eqnarray}
\label{wcbestfit}
C_9^{\mu NP}=-C_{10}^{\mu NP} \in \left[-0.71, -0.35 \right]~~~(\mathrm{at}~1\sigma), \\
C_9^{\mu NP}=-C_{10}^{\mu NP} \in \left[-0.91, -0.18 \right]~~~(\mathrm{at}~2\sigma).
\end{eqnarray}

\subsubsection{Four-quark operators }
\begin{figure}[t!]
\begin{subfigure}[t]{0.5\textwidth}
\centering
\includegraphics[width=5.5cm]{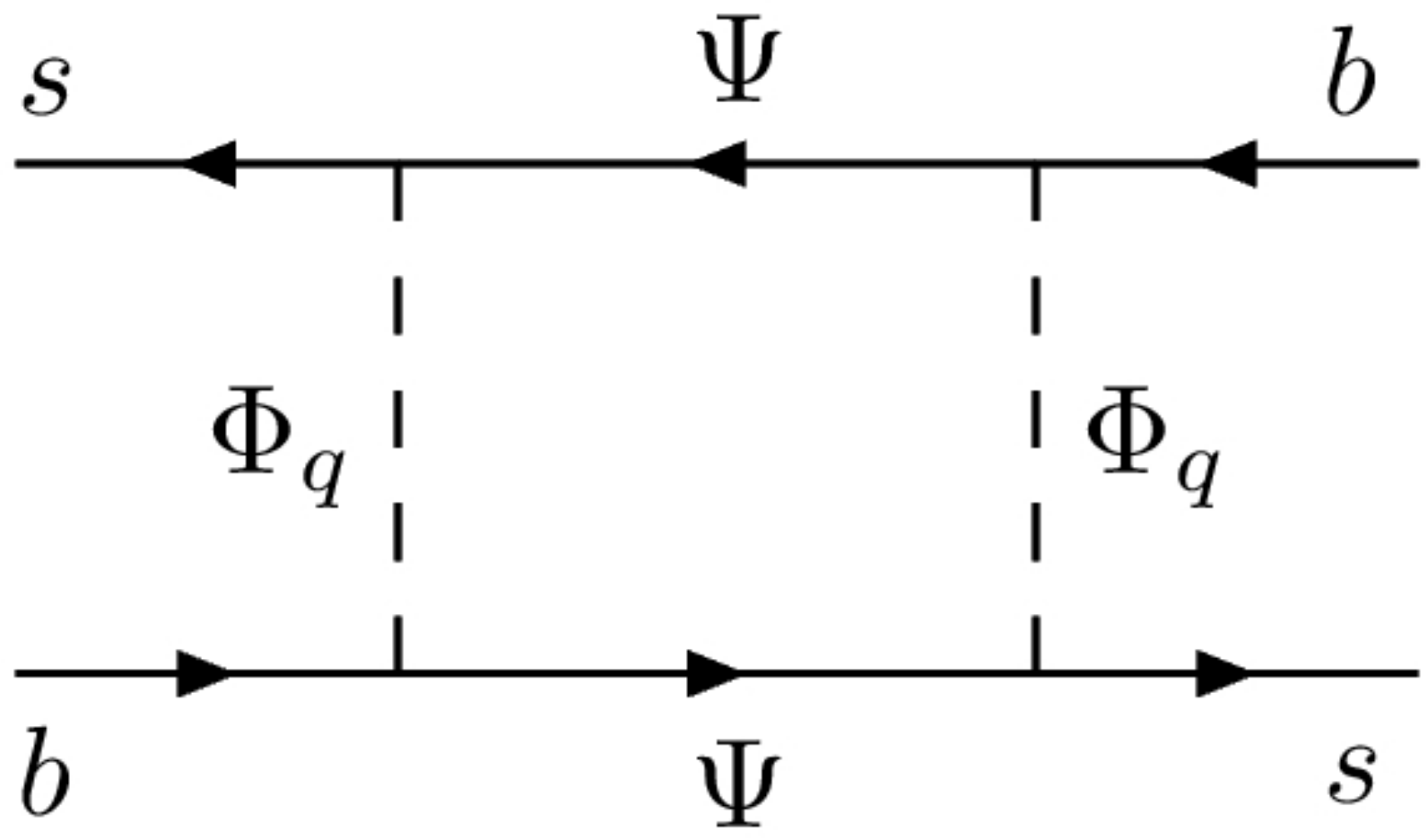}
\end{subfigure}
\begin{subfigure}[t]{0.5\textwidth}
\centering
\includegraphics[width=5.5cm]{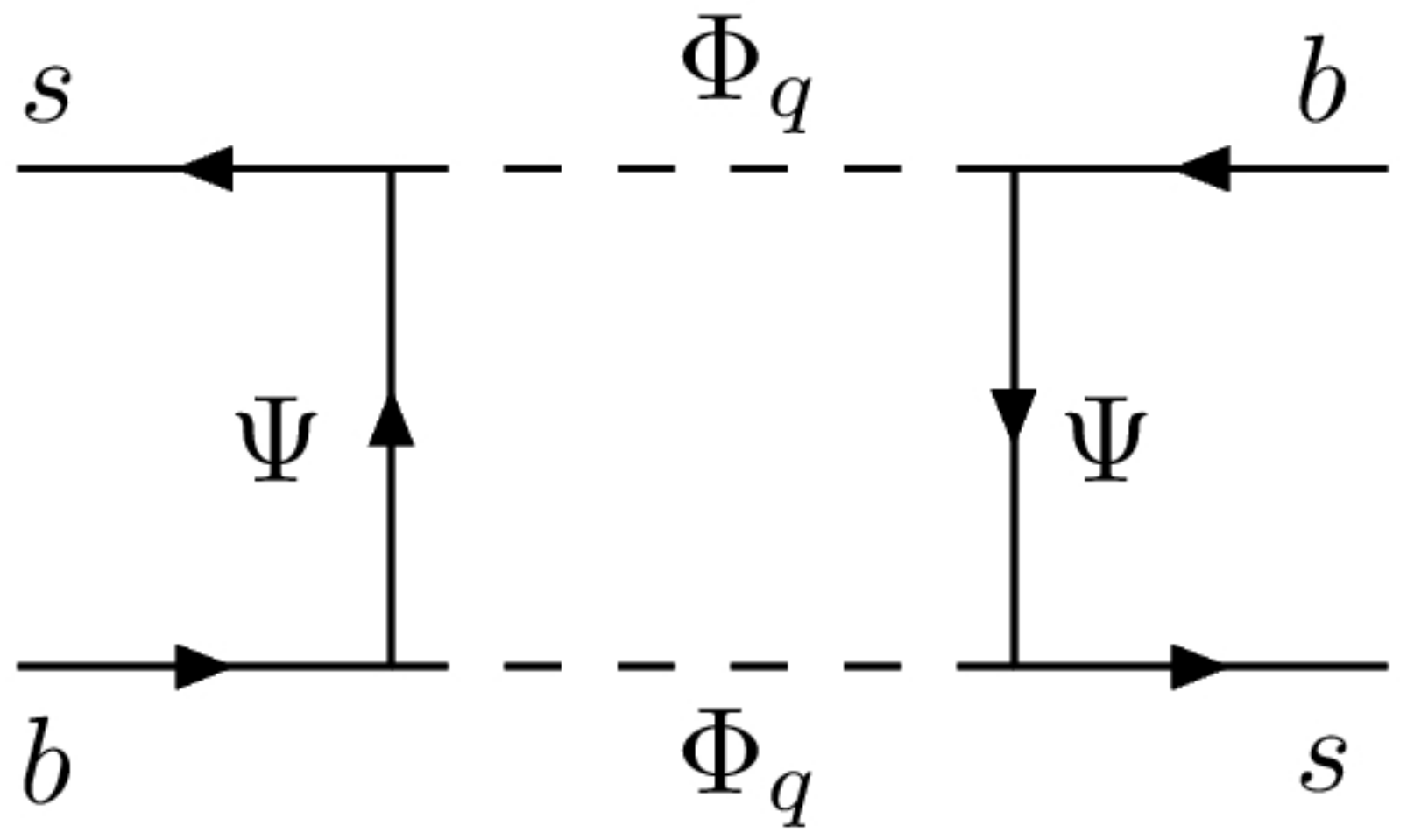}
\end{subfigure}
\caption{Diagrams contributing to $B_s$ mixing}
\label{bsmixing}
\end{figure}

Interactions between four quarks are induced at loop level by diagrams like those in Fig.\@ \ref{bsmixing}. These interactions can lead to meson mixing; in particular, if the process $b \to s \mu \mu$ is present, then inevitably $B_s$ mixing must also be induced. This process can therefore introduce important constraints on the masses and couplings of the new particles. The four quark effective operator induced by the NP is

\begin{equation}
\mathcal{L}_{eff} \supset \frac{K'(x_q)}{M^2_{\Psi}} \frac{\alpha^{q*}_i \alpha^{q}_j \alpha^{q*}_m \alpha^{q}_n}{128 \, \pi^2 } \left[ \left( \overline{Q}^i_L \gamma^{\mu} Q^j_L \right) \left( \overline{Q}^m_L \gamma_{\mu} Q^n_L \right) + \frac{5}{9} \left( \overline{Q}^i_L \gamma^{\mu} \vec{\tau} Q^j_L\right) \cdot \left( \overline{Q}^m_L \gamma_{\mu} \vec{\tau} Q^n_L \right) \right],
\end{equation}
where $K'(x)$ is the first derivative of $K(x)$. The $SU(2)_L$ structure of the effective operator is similar to that of Eqn.\@ \ref{bsllNPeffL} and can again be derived from the discussion in Appendix \ref{SUtwodecomp}.
Projecting the quark doublet along the down components we find that for $B_s$ mixing the relevant operator is 
\begin{equation}
\mathcal{L}_{eff} \supset  \frac{7}{576 \pi^2} \frac{K'(x_q)}{M^2_{\Psi}} \left(\alpha^{q*}_2 \alpha^{q}_3 \right)^2 (\overline{s}_L \gamma^{\mu} b_L) (\overline{s}_L \gamma_{\mu} b_L) + \textrm{ h.c.\@}.
\end{equation}

The Wilson coefficient is easily extracted at high energy $\mu = \Lambda$ where the BSM particles are dynamical fields. We fix $\Lambda =  1$ TeV in what follows. At this energy we have
\bea
C^{bs}_1 (\Lambda) = \frac{7}{576 \pi^2} \frac{K'(x_q)}{M^2_{\Psi}} \left(\alpha^{q*}_2 \alpha^{q}_3 \right)^2 
\eea

In order to place bounds on the parameters of our model, we take into account QCD effects using the results and procedure of  \cite{Buras:2001ra}. Using the anomalous dimension of this work we found that the running of Wilson coefficient from the scale of the New Physics ($\Lambda$) to the scale of the process ($m_b$) is given by $C^{bs}_1 (m_b) = \eta_{VLL} C^{bs}_1 (\Lambda)$ with $\eta_{VLL}=0.78$.
For the evaluation of the relevant matrix element we used the lattice result of \cite{Carrasco:2013zta}.
These lead to a constraint (at 95\% confidence level) on the coefficient 
\begin{equation}
C_1^{bs} (\Lambda) \lesssim \frac{1.8 \times 10^{-5}}{\rm{TeV}^2}
\end{equation}
which translates into
\begin{equation}
\frac{7}{576 \pi^2} \frac{K'(x_q)}{M^2_{\Psi}} \left| \alpha^{q*}_2 \alpha^{q}_3 \right|^2 < 1.8 \times 10^{-5} \left( \frac{1}{\textrm{1 TeV}} \right)^2.
\end{equation}

Thus the measurement of $B_s$ mixing produces a bound on the hadronic couplings involved in the $b \to s \mu \mu$ process, {\em viz.} $\alpha_2^q$ and $\alpha_3^q$. The model can hence accommodate both this bound and the $b \to s \mu \mu$ data if the muonic coupling $\alpha_2^{\ell}$ is sufficiently large. In this respect the model is similar to $Z^{\prime}$ models --- the couplings involved factorize into leptonic couplings and hadronic couplings which can be set independently. This factorization does not occur in leptoquark models.

\subsubsection{$b \to s \gamma$}

\begin{figure}
\begin{center}
\includegraphics[width=5cm]{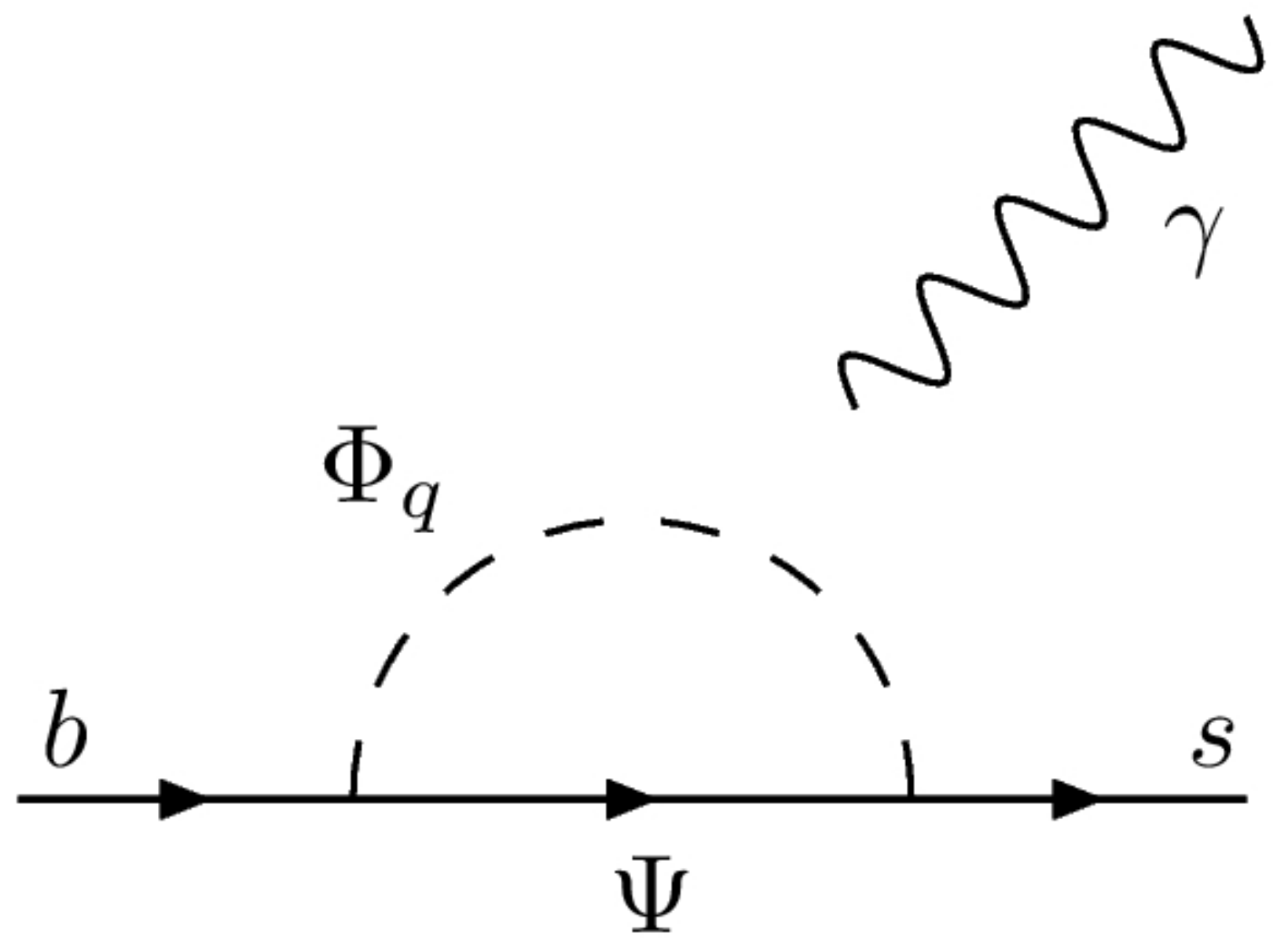}
\end{center}
\caption{Diagram contributing to $b \to s \gamma$. The photon is attached in all possible ways.}
\label{bsgamma}
\end{figure}

The radiative process $b \to s \gamma$ will also be induced by the diagram in Fig.~\ref{bsgamma}. The couplings involved are the same as those for $B_s$ mixing. However, the amplitudes will scale differently with the parameters $\alpha_q$ and $M_q$ between the two processes. Constraints from $b \to s \gamma$ could therefore provide complementary information.

At the mass of the $b$ quark, the process $b \to s \gamma$ is described by the following effective hamiltonian:
\begin{equation*}
{\mathcal H}_{\rm eff} = - 
\frac{4 G_F}{\sqrt{2}}\, (V_{ts}^\ast V_{tb}) \left[ C_7(m_b) {\mathcal O}_7 (m_b) + C_7^{\prime}(m_b) {\mathcal O}_7^{\prime} (m_b)\right],
\end{equation*}
where ${\mathcal O}_7^{(')} = \frac{e}{16\pi^2}  
m_b \left (\bar s \sigma_{\alpha \beta} P_{R(L)} b \right)F^{\alpha \beta}$.

\begin{figure}
\begin{center}
\includegraphics[width=5cm]{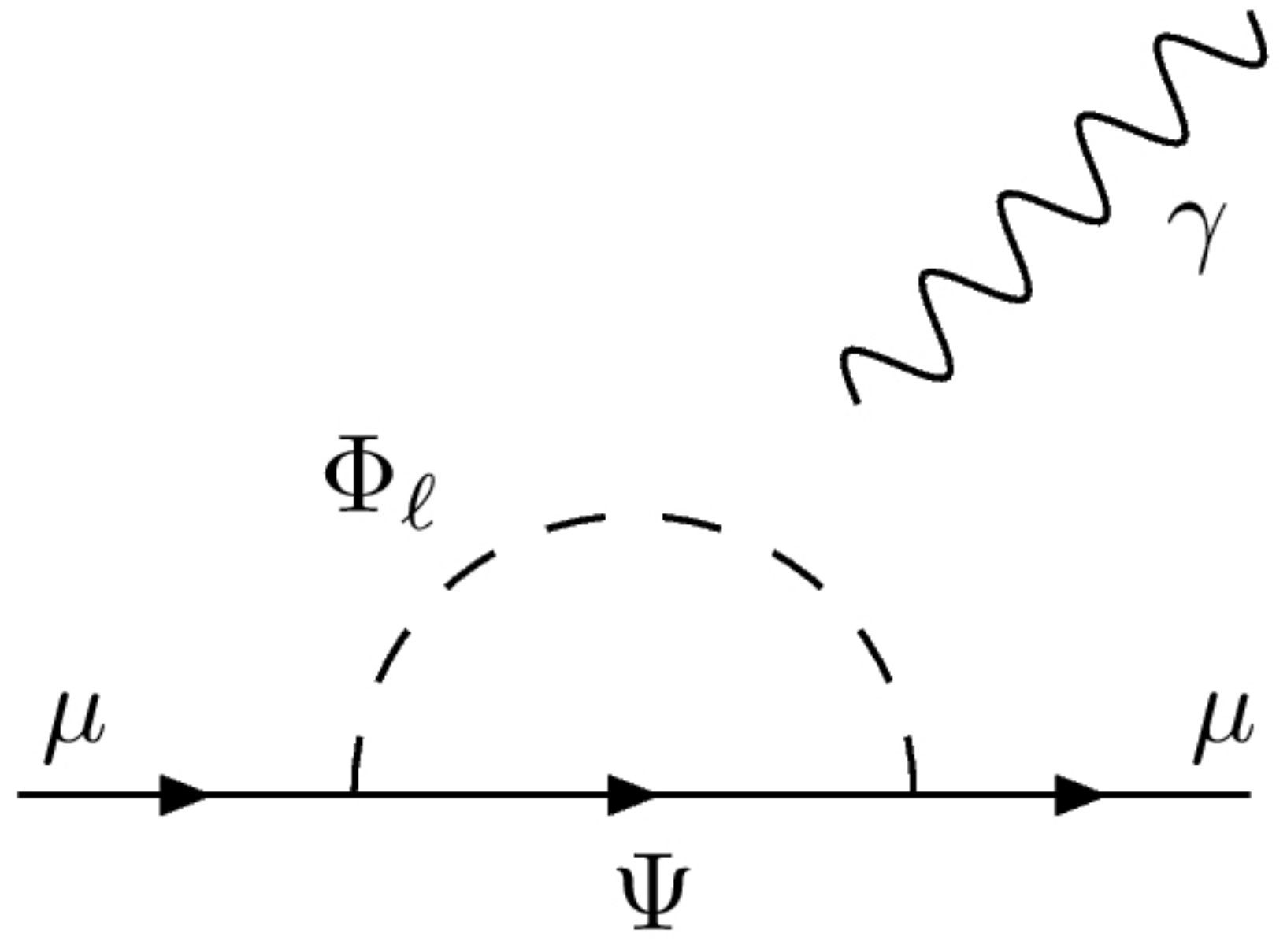}
\end{center}
\caption{Diagram contributing to the anomalous magnetic moment of the muon. The photon is attached in all possible ways.}
\label{mug-2}
\end{figure}
At the matching scale $M$, we get an additional contribution from the NP to the coefficient of the dipole operator;
\begin{equation}
C^{NP}_{7} = \left( \frac{G_F}{\sqrt{2}} V^*_{ts} V_{tb} \right)^{-1} \frac{\alpha_2^{q*} \alpha_3^{q}}{12 M^2_{\Psi}} \left( 3 F_1(x_q) + \frac{2}{x_q} F_1 (x_q^{-1}) \right),
\end{equation}
where $F_1(x)$ is defined as 
\begin{equation}
F_1(x) = \frac{1}{12 (x-1)^4} \left( x^3-6 x^2 +3x+2+6x \log{x} \right) \, .
\end{equation}
The $2 \sigma$ allowed range for this parameter has been fitted recently in \cite{Altmannshofer:2015sma}, giving
\begin{eqnarray}
C_7^{NP} (m_b) \in \left[-0.10, 0.02 \right]~~~(\mathrm{at}~2\sigma).
\end{eqnarray}

\subsubsection{Anomalous magnetic moment of the muon}

Although it is somewhat peripheral to our discussion, let us remark that loops of $\Psi$ and $\Phi_{\ell}$, as shown in Fig.~\ref{mug-2}, generate a 1-loop contribution to the magnetic moment of the muon, which may be able to resolve the long-standing experimental discrepancy therein \cite{Bennett:2006fi}. The NP contribution is given by
\begin{equation}
\Delta a^{NP}_{\mu} =  \frac{\left| \alpha^2_{\ell} \right|^2}{6 \pi^2}\frac{M^2_{\mu}}{M^2_{\Psi}}  \left( 5 F_1(x_{\ell}) + \frac{2}{x_{\ell}} F_1 (x_{\ell}^{-1}) \right),
\end{equation}
which should be compared to the observed discrepancy \cite{Agashe:2014kda}
\begin{equation}
\Delta a_{\mu} =  a^{exp}_{\mu}-a^{SM}_{\mu} = \left( 287 \pm 80 \right) \times 10^{-11}
\end{equation}
As we will show in Section (\ref{plots}), it is possible to fit the anomalous magnetic moment in this model. However, it requires a large value of $\alpha_2^{\ell}$, which is problematic, since it can lead to large corrections to electroweak precision observables at the $Z$-pole.

\subsubsection{$b \to s \nu \overline{\nu}$ processes}
Contributions to $B \to K \nu \overline{\nu}$ and $B \to K^{*} \nu \overline{\nu}$ are expected in the model, due to a diagram similar to Fig.\@ \ref{bsll} with the muons replaced with muon neutrinos (as well as $Z$ penguin diagrams --- see the comment in section \ref{bsllsec}).
A detailed analysis of NP contributions to this process is given in~\cite{Buras:2014fpa}, and we use their results here. Current measurements give bounds on the ratio of total (NP+SM) to SM branching ratios to be
\begin{equation}
\mathcal{R}_K \equiv \frac{BR(B \to K \overline{\nu} \nu)}{BR(B \to K \overline{\nu} \nu)_{SM}} < 4.3
\end{equation}
\begin{equation}
\mathcal{R}_{K^*} \equiv \frac{BR(B \to K^* \overline{\nu} \nu)}{BR(B \to K^* \overline{\nu} \nu)_{SM}} < 4.4
\end{equation}
at 90\% confidence level. An expression for $\mathcal{R}_{K^{(*)}}$ in the presence of NP with couplings to left-handed SM fermions is
\begin{equation}
\mathcal{R}_K=\mathcal{R}_{K^*}=\frac{1}{3} \, \sum_{\ell} \frac{|C_L^{\ell}|^2}{|C_L^{SM}|^2}
\end{equation}
where $C_L^{\ell}$ is the coefficient of the operator 
\begin{equation}
\mathcal{O}_L = \frac{-4G_F}{\sqrt{2}}V_{tb}V^*_{ts} \frac{e^2}{16 \pi^2} \left(\overline{s}\gamma_{\mu} P_L b \right) \left(\overline{\nu}_{\ell} \gamma^{\mu} (1-\gamma_5) \nu_{\ell} \right)
\end{equation}
in the effective Hamiltonian. The SM Wilson coefficient $C_L^{SM}$ is known to be quite accurately $C_L^{SM}=-7.65$.
In the case of our model,
\begin{equation}
C_L^{NP, \, \mu}=\left(\frac{\pi}{\sqrt{2} G_F \alpha V_{tb} V^*_{ts}} \right) \left(-\frac{4}{9} \frac{K(x_q, x_{\ell})}{M_{\Psi}^2} \frac{\alpha_2^q \alpha_3^{q*} |\alpha_2^{\ell}|^2 }{64 \pi^2} \right)
\end{equation}
where the definitions of $K(x,y)$, $x_q$ and $x_{\ell}$ are as in Sec.\@ \ref{bsllsec}. Thus bounds exist on the parameters of the model due to $b \to s \overline{\nu} \nu$ processes:
\begin{equation}
-24.9 < C_L^{NP, \, \mu} < 30.0,
\end{equation}
\begin{equation}
-55.4 < K(x_q, x_{\ell}) \, \alpha_2^q \alpha_3^{q*} |\alpha_2^{\ell}|^2 \left( \frac{M_{\Psi}}{\rm{TeV}} \right)^{-2} < 66.8,
\end{equation}
at 90\% confidence level. However, if we assume that couplings to muons are the dominant leptonic coupling in the model, then we find the relation
\begin{equation}
C_L^{NP}=\frac{2}{7} C_9^{NP,\, \mu}=-\frac{2}{7} C_{10}^{NP,\, \mu}.
\end{equation}
Therefore, for values of the Wilson coefficients required to fit the $b \to s \mu \mu$ anomalies (Eqn.\@ \ref{wcbestfit}), the NP contributions to the branching ratios for the $b \to s \nu \nu$ processes are well below the bounds, adding approximately $5\%$ to the SM values.

\subsection{Direct Searches}
The particles of the three new multiplets, $\Phi_q$, $\Phi_{\ell}$, and $\Psi$, will be directly produced at the LHC if their masses are within kinematic reach. In this subsection, we outline current limits on their masses from direct searches, and identify promising channels to search for them. It will be convenient to label the $SU(2)_L$ components of the multiplets by a superscript denoting their respective electric charges; the full list of NP particles is, therefore, $\Phi_q$=($\Phi_q^{+7/3},\Phi_q^{+4/3},\Phi_q^{+1/3}$), $\Phi_{\ell}$=($\Phi_{\ell}^{+3}$,$\Phi_{\ell}^{+2}$,$\Phi_{\ell}^{+1}$), $\Psi$=($\Psi^{+3}$,$\Psi^{+2}$,$\Psi^{+1}$,$\Psi^{0}$).

The collider phenomenology of the new particles will depend on the mass spectrum. As before, we assume that $M_{\Psi} < M_{\ell} ,M_{q}$, since we require the LP to be neutral. We further assume that $M_{\ell} <M_{q}$, since this minimises contributions to $B_s$ mixing (it also maximises contributions to the muonic $g-2$).
 The three multiplets can, generically, be well-separated in mass, but within each $SU(2)_L$ multiplet there may also be significant mass splittings. For the scalar multiplets, there are tree level mass splittings due to the presence of direct couplings with the Higgs; for the fermion multiplet, there are only small radiative mass splittings between the components.
In the limit that the common mass is much larger than the electroweak scale $v$, the radiative mass splitting between the different charge eigenstates is \cite{Cirelli:2005uq,DelNobile:2009st}
\begin{equation}
\Delta m_{\rm{rad}}=m_{Q+1}-m_Q \approx 166 \rm{MeV} \left( 1+2Q+\frac{2Y}{\cos \theta_W} \right),
\label{massdiff}
\end{equation}
a formula which holds for both scalars and fermions. According to eq.~\ref{massdiff} the lightest particle within the fermion multiplet will be uncharged, as desired.

As the $\Psi$ fermion multiplet has the lowest common mass of all the new states, and due to the $U(1)_\chi$ symmetry within the new sector, the lightest state within the multiplet will be stable.
The small radiative mass splitting means that heavier fermion components will decay to the lightest (neutral) component by emission of one or more soft charged pions or leptons, which will not be energetic enough to be reconstructed in the detector. Thus if any $\Psi$ particle is produced at the LHC, it will appear as missing transverse momentum, similarly to the Wino-like dark matter described in \cite{Cirelli:2014dsa}\footnote{Note that our setup is subtly different from that described in \cite{Cirelli:2014dsa}. There, strong constraints can be put on the mass of the new fermion multiplet from disappearing tracks searches, since the lifetime of a charged fermion decaying to the neutral fermion can be long enough to create a disappearing track in the detector. Here, these searches are not constraining because the lifetime is too short for a track to be visible.}. We thus neglect henceforth the soft undetectable pions or leptons emitted in the decays of heavier components of $\Psi$. Therefore, $\Phi_q$ and $\Phi_{\ell}$ particles (being heavier than the $\Psi$ states) will effectively decay to a SM particle plus missing transverse energy. Furthermore, due to the $U(1)_{\chi}$ symmetry, NP particles will always be pair-produced at the LHC. This means that searches for (R-parity conserving) supersymmetry should be sensitive to $\Phi_q$ and $\Phi_{\ell}$. We will now discuss each of the fields $\Phi_q$, $\Phi_{\ell}$ and $\Psi$ in turn. 

The fermions $\Psi^X$ can be pair produced via a photon or a $W/Z$, or through the decay of $\Phi_{q,\ell}$ . By the arguments above, they will always behave as uncharged weakly interacting particles. Limits can be set on these from mono-x searches, from constraints on the invisible width of the $Z$ boson, and from LEP searches for charginos that are almost degenerate in mass with the neutralinos. A detailed analysis of all of these has been performed in \cite{DiLuzio:2015oha}; the last-mentioned has been found to be the most constraining, implying a bound  of $m_{\Psi}>90$ GeV. 

\begin{figure}
\begin{center}
\includegraphics[width=7cm]{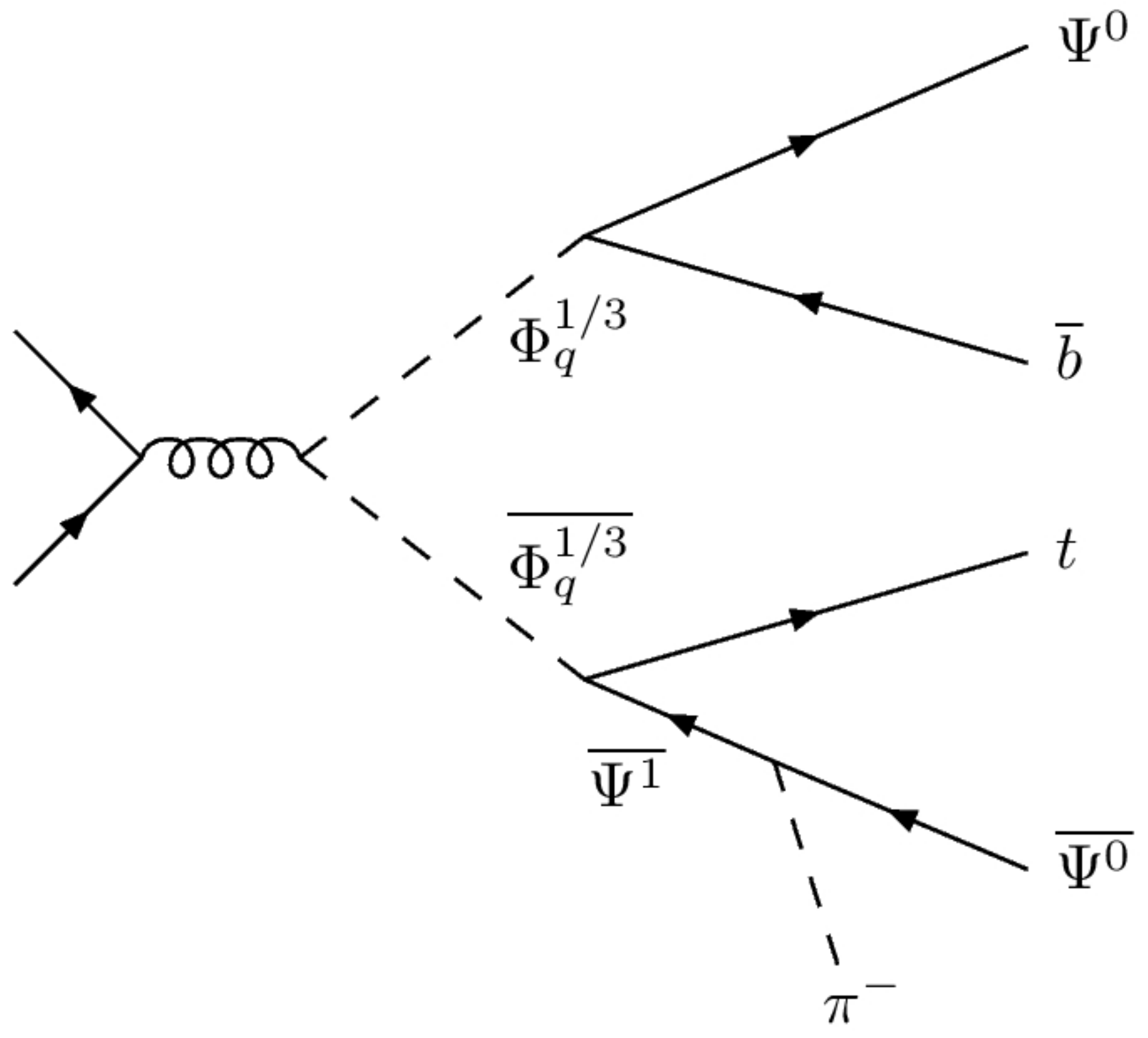}
\end{center}
\caption{Diagram to illustrate the possibility of pair produced $\Phi_q$ particles decaying to two different quarks. Since the pion produced when the $\overline{\Psi^1}$ decays to $\overline{\Psi^0}$ is too soft to be detected, the decay of the $\Phi_q^{1/3}$ appears similar to that of a (anti-)sbottom, whereas the decay of the $\overline{\Phi_q^{1/3}}$ appears similar to that of a stop.}
\label{decaychains}
\end{figure}

Each component ($\Phi_q^{+7/3},\Phi_q^{+4/3},\Phi_q^{+1/3}$) of the coloured scalar multiplet $\Phi_q$ will be strongly pair produced and will decay with a similar signature to that of a squark; \emph{i.e.} to a quark and (either directly or via emission of soft pions or leptons) the stable neutral component of the fermion $\Psi$. Note that there is the possibility that two $\Phi_q$ particles produced each decay to a different flavour of quark. An example decay chain displaying this property is illustrated in Fig.\@ \ref{decaychains}. This complicates the re-interpretation of SUSY search limits.

The flavour considerations discussed earlier only constrain the product of couplings $\alpha_3^q\alpha_2^q$, without constraining their quotient. The strongest constraints from existing LHC searches will hold for situations where one of the couplings $\alpha_2^q$ or $\alpha_3^q$ is much larger than the other, so that the branching ratio to a particular generation dominates. If $\alpha_3^q \gg \alpha_2^q$ the $\Phi_q$ particles will decay like sbottoms and stops, whereas if $\alpha_3^q \ll \alpha_2^q$ they will decay like second-generation squarks. The branching ratio to an up-type quark as opposed to a down-type within a particular generation is determined by the $SU(2)_L$ structure. 

We focus on the case $\alpha_3^q \gg \alpha_2^q$, since this is motivated by flavour considerations, as explained in the next section.
In this limit, one can show that the total branching ratio times cross-section for a pair of $\Phi_q$ particles to be produced and to decay to a pair of tops is $7/8$ of that for direct stop pair production. The most recent limits on direct stop pair production are given in \cite{CMS-PAS-SUS-14-011,Aad:2015pfx}. As a conservative estimate, given that we have the limit $m_{\Psi}>90$ GeV, we can take the limits on direcly pair-produced stops decaying to tops and $90$ GeV neutralinos to apply to our $\Phi_q$. This gives a limit of $M_q \gtrsim 750$ GeV. Likewise, the total branching ratio times cross-section for a pair of $\Phi_q$ particles to be produced and to decay to a pair of $b$ quarks is $7/8$ of that for direct sbottom pair production. Latest limits on direct sbottom pair production are given in \cite{CMS-PAS-SUS-13-018}, and again taking these limits to apply to our $\Phi_q$ particles, we find that $M_q \gtrsim 720$ GeV for a $\Psi$ of mass 90 GeV.

\begin{figure}
\begin{center}
\includegraphics[width=9cm]{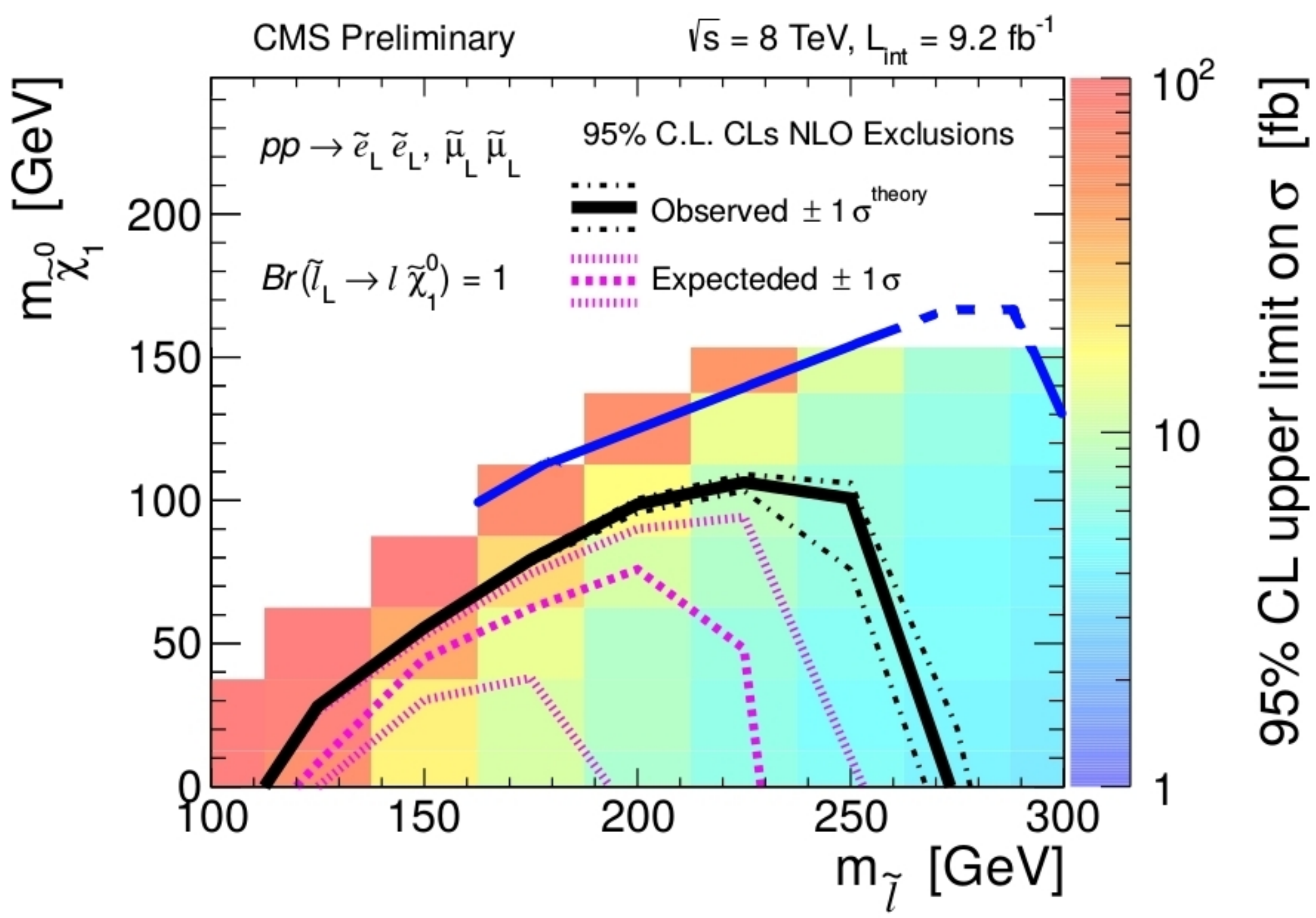}
\end{center}
\caption{Exclusion plot for direct pair production of sleptons decaying to a lepton and a neutralino, taken from \cite{CMS-PAS-SUS-12-022}. The region under the blue line shows the exclusion on our model found by reinterpreting the exclusion plot in terms of direct pair production of $\Phi_{\ell}$s decaying to a lepton and a $\Psi$. For our model the $x$-axis should be taken to mean the mass of the $\Phi_{\ell}$, $M_{\ell}$, while the $y$-axis means the mass of the $\Psi$, $M_{\Psi}$. The dotted part of the blue line is extrapolated.}
\label{exclusion}
\end{figure}

The $SU(2)_L$ components ($\Phi_{\ell}^{+3}$,$\Phi_{\ell}^{+2}$,$\Phi_{\ell}^{+1}$) of the scalar $\Phi_{\ell}$ will, if they have only muonic couplings, always decay to either a muon or a muon neutrino, together with a $\Psi$ particle. So they will sometimes decay in the same way as a smuon in supersymmetry. However, the production cross-sections and branching ratios will differ. Results of recent LHC slepton searches are given in \cite{CMS-PAS-SUS-12-022,Aad:2014vma}. These rule out left-handed smuons, pair produced directly via a $W/Z/\gamma$ and decaying to a muon and a neutralino, up to a maximum mass of roughly 300 GeV (for a massless neutralino). We used Feynrules \cite{Alloul:2013bka,Degrande:2011ua} and Madgraph5\_aMC$@$NLO \cite{Alwall:2014hca} to calculate electroweak (EW) pair production cross-sections of the $\Phi_{\ell}$ particles, and then multiplied by the branching ratios in order to reinterpret the limits on cross-section given in \cite{CMS-PAS-SUS-12-022}. The CMS limit plot, with our model superimposed in blue, is shown in figure~\ref{exclusion}. If the mass of the $\Psi$ particle is greater than about 150 GeV, there are no bounds on the mass of the $\Phi_{\ell}$ (other than the assumption that its mass is greater than that of $\Psi$).

\subsection{Parameter space plots}
\label{plots}
In this subsection we show allowed regions in the parameter space of the model considering the observables described above; $b \to s \ell \ell$, $B \to X_s \gamma$, $B_s$ meson mixing and the anomalous magnetic moment of the muon. The relevant parameters entering the expressions of these observables are the masses of the new states ($M_{\Psi}, M_q$ and $M_{\ell}$) as well as the coupling to muons $\alpha^{\ell}_2$ and the combination $\alpha_2^{q*} \alpha_3^{q}$. Without loss of generality, we can re-define the BSM fields to make these parameters real.

\begin{figure}
\begin{center}
\includegraphics[scale=1.0]{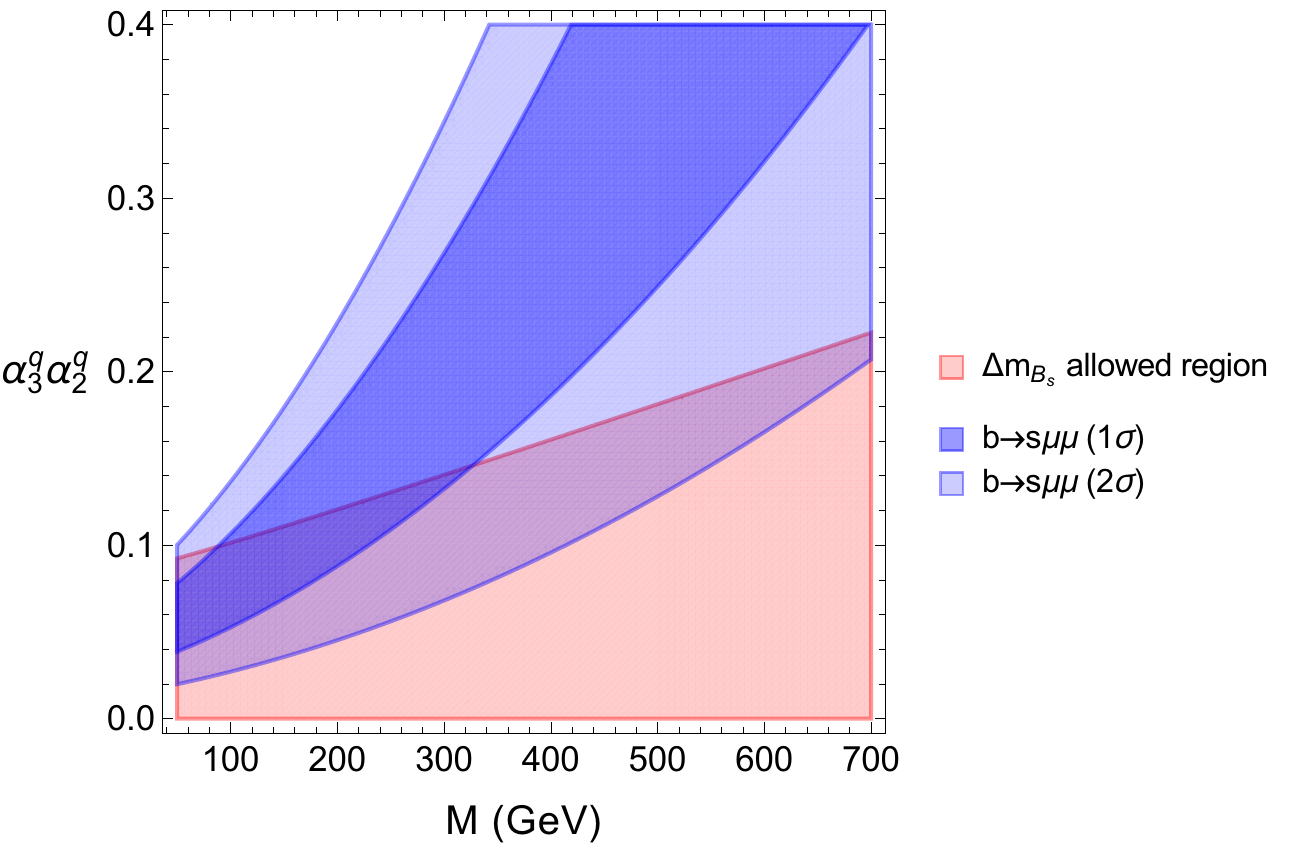}
\end{center}
\caption{Parameter space plot for $\alpha_{2}^{\ell}=1.2$, and with the masses of the three fields given by $M_{\Psi} =M,
M_{\ell}=M +  \textrm{200 GeV}, M_{q} = M + \textrm{700 GeV}$. For this value of $\alpha_2^{\ell}$, it is not possible to explain the anomalous magnetic moment of the muon whilst fitting the other constraints.}
\label{plot2}
\end{figure}

In order to fit the $b \to s \ell \ell$ anomalies without being in disagreement with the measured $B_s$ mixing rate, the muonic coupling $\alpha_2^{\ell}$ must be rather large. In order to have an idea of the typical values of the parameters needed in our model, in Fig.\@ \ref{plot2} we show parameter space regions assuming that  $\alpha^{\ell}_2 = 1.2$ while parametrizing the masses in terms of one single scale $M$ assuming the following hierarchy $M_{\Psi} =M, ~M_{\ell}=M +  \textrm{200 GeV},~ M_{q} = M + \textrm{700 GeV}$. In this way we are left with two parameters only $(M$ and $\alpha_2^{q} \alpha_3^{q} )$. The $B \to X_s \gamma$ allowed region is not shown because it yields weaker constraints than $B_s$ mixing does. For this hierarchy of masses, the only relevant direct production constraint is the bound on the mass of $\Psi$, $M_{\Psi} > 90$ GeV. There is an overlap between the allowed $B_s$ mixing region and the $1 \sigma$ preferred region for the $b \to s \ell \ell$ measurements --- so with these parameters, the model can fit the $b \to s \ell \ell$ anomalies. The value of $\alpha_2^{\ell}$ can be further lowered to be $\lesssim 1$, in this case the values of $M_{\Psi}, M_{\ell}$ and $M_q$ are close to present bounds coming from direct searches. For example we verified that a fit to the data with 
$\alpha_2^{\ell} \approx 0.8$ could be achieved when $M_{\Psi}=150$ GeV, $M_{\ell}=200$ GeV and $M_q=800$ GeV.

\begin{figure}
\begin{center}
\includegraphics[scale=1.0]{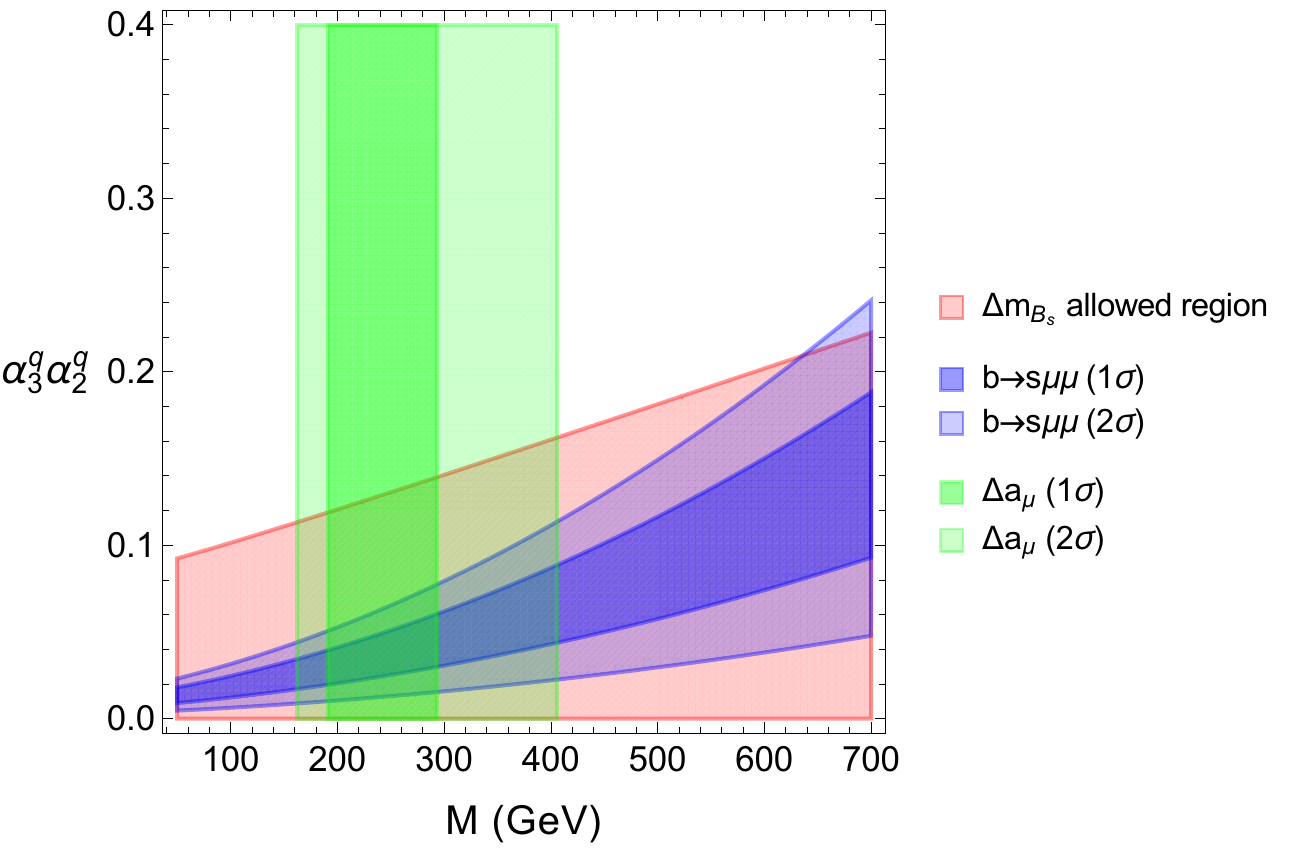}
\end{center}
\caption{Parameter space plot for $\alpha_{2}^{\ell}=2.5$, and with the masses of the three fields given by $M_{\Psi} =M,
M_{\ell}=M +  \textrm{200 GeV}, M_{q} = M + \textrm{700 GeV}$. With this large value of $\alpha_{2}^{\ell}$ there is an overlap between the regions that fit the $B$ anomalies (in blue), and the anomalous magnetic moment of the muon (in green). }
\label{plot1}
\end{figure}

However, if we also wish to fit the anomalous magnetic moment of the muon, the muonic coupling $\alpha_2^{\ell}$ must be larger. We show in Fig.\@ \ref{plot1} the relevant parameter space regions when this coupling is set to $\alpha_2^{\ell}=2.5$, with the same hierarchy of masses as before. If we want to take this explanation of the $(g-2)_{\mu}$ anomaly seriously, then we should consider possible bounds from the shift of the EW gauge couplings $Z\mu \mu, Z \nu_{\mu} \overline{\nu}_{\mu}$  and $W^+ \mu \overline{\nu}_{\mu}$ (see also the discussion in section IV of \cite{Belanger:2015nma}). The corrections are non-universal and so a global fit to EW data is required to establish the precise constraints on the couplings. Though such a fit is beyond the scope of our work, na\"{\i}ve arguments suggest that $O(1)$ values of $\alpha_2^{\ell}$ are not problematic.

\section{Flavour symmetries}
\label{flavour}
In this section we establish a possible connection between the flavour violation present in the SM and in the NP sector.

In the SM and in the limit of vanishing Yukawa couplings, the largest group of unitary field transformations that commutes with the gauge group and leaves invariant the kinetic terms is $U(3)^5 \times U(1)_H$. 
Adopting notation similar to \cite{D'Ambrosio:2002ex} we can decompose this group in the following way;
\begin{equation}
\nonumber
\mathcal{G}_K \equiv  SU(3)^3_q \times SU(3)^2_{\ell} \times U(1)_B \times U(1)_L \times U(1)_Y \times U(1)_{PQ} \times U(1)_{E_R} \times U(1)_H,
\end{equation}
with 
\begin{eqnarray}
SU(3)^3_q &=& SU(3)_{Q_L} \times SU(3)_{U_R} \times SU(3)_{D_R} \\
SU(3)^2_{\ell} &=& SU(3)_{L_L} \times SU(3)_{E_R}.
\end{eqnarray}
The $U(1)$ factors can be identified with the baryon $(B)$ and lepton $(L)$ numbers, the hypercharge $(Y)$, a transformation $(PQ)$ acting non trivially and in the same way only on $D_R$ and $E_R$, and finally a universal rotation for the fields $E_R$ and a $U(1)$ global symmetry associated to the Higgs doublet.

We would like now to make connections with the flavour structure of the SM and the possible effects coming from NP.
In order to do that a first step is to identify \textit{(i)}  a \text{flavour symmetry} and \textit{(ii)} a set of \textit{irreducible} symmetry-breaking terms. 
The flavour symmetry group  $\mathcal{G}_F \supset \mathcal{G}_K$ has to be broken in order to reproduce the observed pattern of fermion masses and mixing. In order to do that a set of 
symmetry-breaking spurions are introduced to formally restore the symmetry $\mathcal{G}_F$. 

We will now consider 3 explicit examples and we will focus on the quark sector.

\begin{enumerate}
\item $\mathcal{G}_F=U(3)^3_q$

This is the case of Minimal Flavour Violation \cite{D'Ambrosio:2002ex}. The spurion fields are the three Yukawa couplings
\begin{equation}
Y_U \sim (3,\overline{3},1) \qquad Y_D \sim (3,1,\overline{3}),
\end{equation}
where the quantum numbers are specified with respect to the direct product of groups $SU(3)_{Q_L} \times SU(3)_{U_R} \times SU(3)_{D_R}$.
\item $\mathcal{G}_F=U(2)^3_q$

This is the flavour symmetry of the quark sector if only the Yukawa couplings $y_t$ and $y_b$ are non-vanishing. So to a good level this is an approximate symmetry of the SM.
Recent works \cite{Barbieri:2011ci,Barbieri:2012uh,Barbieri:2014tja} considered the following set of irreducible spurions ;
\begin{equation}
\Delta_u \sim (2,\overline{2},1), \qquad \Delta_d \sim (2, 1,\overline{2}), \qquad \mathcal{V} \sim (2,1,1),
\end{equation}
where the quantum numbers are specified with respect to the direct product of groups $SU(2)_{Q_L} \times SU(2)_{U_R} \times SU(2)_{D_R}$.

\item $\mathcal{G}_F=U(1)^9$

This case mimics partial compositeness. The irreducible spurions are connected to the Yukawa couplings in the following way;
\begin{equation}
(Y_U)_{ij} \sim \epsilon^q_{i} \epsilon^u_{j}, \qquad (Y_D)_{ij} \sim \epsilon^q_{i} \epsilon^d_{j}.
\end{equation}
\end{enumerate}

With these specific cases in mind we are now ready to discuss flavour violation induced by operators of the form $\alpha^q_i \, \overline{\Psi} Q^i_L \Phi$, $\alpha^u_i \, \overline{\Psi} U^i_R \Phi$ and $\alpha^d_i \, \overline{\Psi} D^i_R \Phi$. 
These operators break the flavour symmetry and in order to restore it we could assume that the vectors $\alpha^F$ are again spurions with definite transformation rules under the flavour symmetry.  We could now \textit{assume} minimality of flavour violation
in the following sense: the $\alpha^F_i$ can be expressed using the irreducible spurions used to construct the SM Yukawa couplings. 
Following this procedure we obtain the following results. 

\begin{enumerate}
\item $\mathcal{G}_F=U(3)^3_q$

To recover flavour invariance the $\alpha^F$ have to transform in the following way;
\begin{equation}
\alpha^q \sim (\overline{3},1,1), \qquad \alpha^u \sim (1,\overline{3},1), \qquad \alpha^d \sim (1,1,\overline{3}).
\end{equation}
However, it can be proved using triality properties of the $SU(3)$ irreps that tensor products of $Y_U,Y^{\dagger}_U,Y_D,Y^{\dagger}_D$ can never give rise to any of the $\alpha^F$. Furthermore, the structure of the Yukawas cannot be reproduced by combinations of the $\alpha$s, since this can only lead to Yukawa matrices of rank 1, meaning two of the up-type quarks and two of the down-type quarks would have zero mass. We therefore conclude that this structure cannot work.

\item $\mathcal{G}_F=U(2)^3_q$

To recover flavour invariance the $\alpha^F$ have to transform in the following way;
\begin{eqnarray}
(\alpha^q_1,\alpha^q_2) \sim (\overline{2},1,1), & \qquad & \alpha^q_3 \sim (1,1,1), \\
(\alpha^u_1,\alpha^u_2) \sim (1,\overline{2},1), & \qquad & \alpha^u_3 \sim (1,1,1), \\
(\alpha^d_1,\alpha^d_2) \sim (1,1,\overline{2}), & \qquad & \alpha^d_3 \sim (1,1,1). 
\end{eqnarray}
Atleading order in the number of spurion fields we have that  
\begin{eqnarray}
(\alpha^q_1,\alpha^q_2)_i = a^q \, \mathcal{V}^{\dagger} _i, \\
(\alpha^u_1,\alpha^u_2)_i = a^u \, (\mathcal{V}^{\dagger} \Delta_u)_i , \\
(\alpha^d_1,\alpha^d_2)_i = a^d \, (\mathcal{V}^{\dagger} \Delta_d)_i ,
\end{eqnarray}
while $\alpha^F_3=b^F$ with $a^F, b^F$ order one numbers and $i=1,2$. Doing a spurion analysis with this setup, we find that in the basis in which $Q_L^i = \left(V^{\dagger ij}_{CKM}u_j, d^i\right)^T$ for $i=1,2$ (generation index) and $Q_L^3=\left(e^{i\Phi_t}t_L,b_L\right)^T$, the coupling vector in the quark sector $\alpha_q$ is found to be

\begin{equation}
\vec{\alpha_q}=\left(y_{12}V_{td}, y_{12}V_{ts}, y_3 \right),
\end{equation}

where $y_{12}$ and $y_3$ are each (generally complex) numbers of $O(1)$.

This is very similar to the case of Partial Compositeness, except that there is slightly less freedom in the couplings; the ratio of $\alpha^q_2$ to $\alpha^q_1$ is exactly fixed (rather than fixed up to a factor of $O(1)$). The natural size of $\alpha_3^q \alpha_2^q$ is $O(|V_{ts}|)\approx \lambda^2$. This ties in well with the allowed regions in Figs.\@ \ref{plot2} and \ref{plot1}. 

\item $\mathcal{G}_F=U(1)^9$

It is easy to show that link between the $\alpha^F_i$ and $\epsilon^F_i$  is simply given by $\alpha^F_i = c^F_i \epsilon^F_i$ where $c^F_i$ are order one numbers. 
\end{enumerate}

Thus, a number of patterns of flavour symmetry breaking can give rise to the hierarchical structure and alignment with the SM Yukawa couplings that is needed to explain the anomalies. In all cases, the pattern of couplings is similar to that arising in models of partial compositeness (see for example \cite{KerenZur:2012fr}). 
Although our phenomenological analysis in the previous section was done with more restrictions on the couplings, we can be sure that a partial compositeness-like flavour structure in the quark sector is phenomenologically viable, since a full analysis of existing bounds on this setup has recently been performed in \cite{Gripaios:2014tna}.

An analogous discussion could be repeated in the leptonic sector, starting from the global symmetry $U(3)_{L_L} \times U(3)_{E_R}$ of the gauge kinetic terms associated with left- and right-handed leptons. However, a complete understanding of the leptonic flavour requires the knowledge of the mechanism that generates neutrino masses. Here we remain agnostic as to the possible flavour orientation of the spurions $\alpha^{\ell}$ and $\alpha^e$. For recent works that consider possible links between the flavour violation in the neutrino sector and the physics of LFV transitions in $b \to s \ell \ell'$, we refer the reader to~\cite{Glashow:2014iga,Boucenna:2015raa,Varzielas:2015iva,Lee:2015qra,Guadagnoli:2015nra}.

\section{Conclusions}
\label{conclusions}
We have presented renormalizable extensions of the SM that can explain several  anomalies observed in $B$-meson decays. Renormalizability (which amounts to the assumption that further NP is heavy and decoupled) allows us to introduce deviations from the SM, coming from NP, in a controlled way. This is almost a {\em sine qua non}, 
given that we observe just a handful of anomalies in data, while many thousands of other observations agree with the SM. 

We have surveyed the possible NP fields that allow for a coupling to linear combinations of left-handed SM fermions, since this generates, at one-loop, an operator of the form $\alpha^q_i \alpha^q_j \alpha^\ell_k \alpha^\ell_l Q^i_L \gamma^\mu Q^j_L L^k_L \gamma_\mu L^l_L$. Coupled with the plausible assumption that the linear flavour-violating spurions, $\alpha^{q,\ell}$ are roughly aligned with the Yukawa couplings of the SM, we end up with a good fit to the anomalies, without contradicting other data.

As a spectacular example of the control that renormalizability brings, the models that we identify feature 3 accidental global symmetries, corresponding to conservation of (generalized) baryon and lepton numbers and to a `NP number'. The consequences of these accidental symmetries are manifold. Not only is the proton stabilized, but also all other baryon- and lepton-number violating processes ({\em e.g.} neutron-antineutron oscillations, $\mu \rightarrow e \nu \nu$), many of which are strongly constrained, are forbidden automatically. The NP number leads to a generic suppression of NP flavour-violating processes, since these can only occur at loop level. Yet another advantage of the models is that NP can only couple to left-handed SM fermions at the renormalizable level, meaning that contributions to processes requiring a helicity flip, such as $\mu \rightarrow e \gamma$, are further suppressed.

The accidental symmetries of the models, while sufficient to prevent many dangerous processes, are quite different from the accidental symmetries of the SM, namely baryon and individual lepton family numbers. This is a crucial feature, since it allows us to have large violations of lepton universality. This is precisely what is needed to fit the anomalies. 

The models are not panace\ae, in that there is a further anomaly in $B$ physics that cannot be explained, arising in decays to $D^{(*)} \tau \nu$ \cite{Lees:2013uzd,Lees:2012xj,Huschle:2015rga}. But it seems hard to explain this anomaly in any NP model, for the simple reason that the SM contribution, with which it needs to be comparable, is so large (being a tree-level effect with minimal Cabibbo suppression)\footnote{However it is not impossible, see for example \cite{Fajfer:2012jt,Alonso:2015sja,Freytsis:2015qca}.}.

The main weakness of the models is arguably that they require a rather large value of the coupling $\alpha_2^{\ell}$ in order to explain the anomalies. While this coupling can be rather smaller than other couplings in the flavour sector ({\em i.e.} the top quark Yukawa coupling), some readers may be alarmed that such a large coupling should appear in the
light lepton sector, where (at least in the SM) all other couplings are small.
It is important to note, however, that not only does this coupling not cause phenomenological problems {\em per se},\footnote{In fact, as shown in \cite{Giudice:2011ak}, one can even put $O(1)$ couplings among the light quark generations without necessarily getting into trouble.} but also that, provided that it is suitably aligned, it does not lead to large flavour violations in the light leptons via renormalization group flow. This follows immediately from the observation that there exists a basis in which the SM leptonic Yukawa couplings are diagonal. Nevertheless, the necessary alignment is aesthetically disturbing; we have shown that it is plausible from the point of view of flavour symmetries, but it would be nice to have an explicit model of flavour in which it is realised dynamically.
\section*{Acknowledgments}
This work has been partially supported by the Galileo Galilei Institute for Theoretical Physics, STFC grant ST/L000385/1, and King’s College, Cambridge. We thank members of the Cambridge SUSY Working Group for discussions. 

\appendix
\section{SU$\mathbf{(2)_{\it L}}$ decompositions}
\label{SUtwodecomp}

By denoting the generators in the fundamental representation of SU$(2)_L$ as $T^a = \sigma^a/2$ 
(with $\sigma^a$ being the Pauli matrices and $a=1,2,3$), 
we define their action on the $(2j+1)$-dimensional completely symmetric tensor $\chi_{i_{1}i_{2} \ldots i_{2j}}$ 
($i_{1}, i_{2}, \ldots,  i_{2j} =1,2$) as
\begin{equation}
\delta^a (\chi_{i_{1}i_{2} \ldots i_{2j}}) = 
T^a_{i_1 k} \, \chi_{k i_{2} \ldots i_{2j}} 
+ T^a_{i_2 k} \, \chi_{i_{1} k \ldots i_{2j}} 
+ \ldots 
+ T^a_{i_{2j} k} \, \chi_{i_{1} i_{1} \ldots k} 
\, . 
\end{equation}
In general, we arrive at the following embedding of the properly normalized electric charge eigenstates: 
\begin{equation}
\begin{array}{l}
\chi_{11 \ldots 1} = \frac{1}{\sqrt{B_{2j,0}}} \chi^{Q} \\
\chi_{11 \ldots 2} = \frac{1}{\sqrt{B_{2j,1}}} \chi^{Q-1} \\ 
\vdots \\ 
\chi_{12 \ldots 2} = \frac{1}{\sqrt{B_{2j,2j-1}}} \chi^{Q-2j+1} \\ 
\chi_{22 \ldots 2} = \frac{1}{\sqrt{B_{2j,2j}}} \chi^{Q-2j} \, ,
\end{array}
\end{equation}
where the superscripts denote the electric charge of the field, $B_{n,k}$ is the binomial factor 
$B_{n,k} = \frac{n!}{k! (n-k)!}$ and the normalization of the states is such that 
\begin{equation}
\chi^{*i_{1}i_{2} \ldots i_{2j}} \chi_{i_{1}i_{2} \ldots i_{2j}} = |\chi^{j}|^2 + |\chi^{j-1}|^2 + 
\ldots + |\chi^{-j+1}|^2 + |\chi^{-j}|^2 \, . 
\end{equation}

In the following, we provide the SU$(2)_L$ decomposition for the BSM fields ($\Psi,\Phi_q$ and $\Phi_{\ell}$ ) introduced in the Model A (\ref{ModelA}):

\begin{equation}
\begin{array}{ccl}
\Psi &=& (1,4,\tfrac{3}{2}) \\
\Psi_{111} &=& \Psi^{3} \nonumber \\
\Psi_{112} &=& \tfrac{1}{\sqrt{3}} \Psi^{2}  \nonumber \\
\Psi_{122} &=& \tfrac{1}{\sqrt{3}} \Psi^{1} \\
\Psi_{222} &=& \Psi^{0} \nonumber 
\end{array}
\qquad
\qquad
\begin{array}{ccl}
\Phi_q &=& (\overline{3},3,\tfrac{4}{3}) \\
(\Phi_q)_{11} &=&\Phi_q^{7/3} \nonumber \\
(\Phi_q)_{12} &=& \tfrac{1}{\sqrt{2}} \Phi_q^{4/3}  \nonumber \\
(\Phi_q)_{22} &=& \Phi_q^{1/3} 
\end{array}
\qquad
\qquad
\begin{array}{ccl}
\Phi_\ell &=& (1,3,2) \\
(\Phi_\ell)_{11} &=&\Phi_\ell^{3} \nonumber \\
(\Phi_\ell)_{12} &=& \tfrac{1}{\sqrt{2}} \Phi_\ell^{2}  \nonumber \\
(\Phi_\ell)_{22} &=& \Phi_\ell^{1} 
\end{array}
\end{equation}
The relevant linear interactions (\ref{lin}) introduced in  Model A can be rewritten in the following way
\begin{eqnarray}
\mathcal{L}_{\textrm{lin}} &=& \alpha^q_i \, \overline{\Psi}_R  Q^i_L \Phi_q  + \alpha^{\ell}_i \, \overline{\Psi}_R L^i_L \Phi_{\ell} +  \textrm{ h.c.}  \\
&=&
\alpha^q_i \, (\overline{\Psi}_R)^{k_1 k_2 k_3}  (Q^i_L)_{k_1} (\Phi_q)_{k_2k_3}  + \alpha^{\ell}_i \,  (\overline{\Psi}_R)^{k_1k_2k_3}   (L^i_L)_{k_1} (\Phi_{\ell})_{k_2k_3} +  \textrm{ h.c.}, 
\end{eqnarray}
where $k_1,k_2,k_3= \{ 1,2 \}$ are $SU(2)_L$ fundamental indices. More explicitly we get
\begin{eqnarray}
\label{SU2decomplagrangian}
\mathcal{L}_{\textrm{lin}} &=& \alpha^q_i \left(\overline{\Psi}^{-3}_R \Phi^{7/3}_q + \sqrt{\frac{2}{3}} \, \overline{\Psi}^{-2}_R \Phi^{4/3}_q + \sqrt{\frac{1}{3}} \, \overline{\Psi}^{-1}_R \Phi^{1/3}_q \right) u^i_L +  \textrm{ h.c.} \\
&+& \alpha^q_i \left( \sqrt{\frac{1}{3}} \, \overline{\Psi}^{-2}_R \Phi^{7/3}_q + \sqrt{\frac{2}{3}} \, \overline{\Psi}^{-1}_R \Phi^{4/3}_q + \overline{\Psi}^{0}_R \Phi^{1/3}_q \right) d^i_L +  \textrm{ h.c.} \\
&+& \alpha^{\ell}_i \left(\overline{\Psi}^{-3}_R \Phi^{3}_\ell + \sqrt{\frac{2}{3}} \, \overline{\Psi}^{-2}_R \Phi^{2}_\ell + \sqrt{\frac{1}{3}} \, \overline{\Psi}^{-1}_R \Phi^{1}_\ell \right) \nu^i_L +  \textrm{ h.c.} \\
&+& \alpha^\ell_i \left( \sqrt{\frac{1}{3}} \, \overline{\Psi}^{-2}_R \Phi^{3}_\ell + \sqrt{\frac{2}{3}} \, \overline{\Psi}^{-1}_R \Phi^{2}_\ell + \overline{\Psi}^{0}_R \Phi^{1}_\ell \right) e^i_L +  \textrm{ h.c.} 
\end{eqnarray}

\section{Photon and $Z$-boson mediated contributions}
\label{penguins}

An explicit calculation of the photon penguin diagrams leads to the following (lepton flavour universal) contribution to the Wilson coefficient of the leptonic vector current at low energy;
\begin{eqnarray}
C^{\gamma}_9 &=&  \left( \frac{4 G_F}{\sqrt{2}} V^*_{ts} V_{tb} \frac{\alpha}{4 \pi}\right)^{-1} \frac{\alpha^{q*}_2 \alpha^{q}_3 } {M^2_{\Psi}} \frac{\alpha}{4 \pi} \frac{2}{27} x_q^{-1} f_{\gamma}(x_q^{-1}),
\end{eqnarray} 
where 
\begin{equation}
f_{\gamma} (x) = - \frac{2 (38-79 x + 47 x^2) }{9 (1-x)^3} - \frac{4 (4-6 x + 3 x^3) }{3 (1-x)^4} \log x,
\end{equation}
and where, as before, $x_q \equiv \frac{M_q^2}{m_{\Psi}^2}$. This function is normalised in such a way that $f_{\gamma} (1) =1$.

In our model, the contribution of the $Z$-boson mediated penguin diagram is suppressed compared to that of the photon mediated one. Indeed the $Z$-boson exchange is enhanced only in diagrams containing a source of explicit $SU(2)_L$ breaking\footnote{This general argument has been given in the context of $Z$-penguin contributions in the MSSM in \cite{Lunghi:1999uk} and in \cite{Bertolini:1990if}.}; such diagrams are not present in our case. When there is no explicit $SU(2)_L$  breaking, the contribution from the $Z$-boson penguin diagram is suppressed by a factor $\frac{m^2_B} {M^2_Z} \sim 3 \times 10^{-3}$ compared to the photon one; this factor is given simply by the ratio of the propagators in the two cases.

Neglecting the $Z$-boson contribution, we now quantitatively show that the photon contribution is suppressed at the percent level when compared to the one in Eq. (\ref{boxbsll}). 
Taking the ratio of the Wilson coefficients generated through a photon penguin and the NP box diagram respectively, we get
\begin{equation}
\frac{C^{\gamma}_9}{C^{\mu NP}_9} = \frac{\frac{\alpha}{4 \pi} \frac{2}{27} x_q^{-1} f_{\gamma}(x_q^{-1}) }{\frac{7}{576 \pi^2} K({x_q,x_{\ell}}) \left| \alpha^{\ell}_2 \right|^2}
= 3.5 \times 10^{-2} \frac{ x_q^{-1} f_{\gamma}(x_q^{-1}) }{K({x_q,x_{\ell}}) \left| \alpha^{\ell}_2 \right|^2}.
\end{equation}
Taking as a reference the benchmark defined in Figure 8, namely  $\alpha^{\ell}=2.5, M_{\Psi} =M,
M_{\ell}=M +  \textrm{200 GeV}, M_{q} = M + \textrm{700 GeV}$ we find that
\begin{equation}
\left| \frac{C^{\gamma}_9}{C^{\mu NP}_9} \right| < 3.9 \times 10^{-2} \qquad \textrm{for } M > 150 \textrm{ GeV.}
\end{equation}

\bibliographystyle{JHEP}
\bibliography{references}
\end{document}